\documentclass[sigconf]{acmart}
\usepackage{caption}
\usepackage{graphicx}
\usepackage{float} 
\usepackage{subcaption}
\usepackage{pifont}
\usepackage[linesnumbered,ruled,vlined]{algorithm2e}
\usepackage{url}
\usepackage{multirow}
\usepackage{hyperref}
\usepackage{xcolor}
\AtBeginDocument{%
}

\copyrightyear{2025}
\acmYear{2025}
\setcopyright{acmlicensed}
\acmConference[WWW '25] {Proceedings of the ACM Web Conference 2025}{April 28--May 2, 2025}{Sydney, NSW, Australia.}
\acmBooktitle{Proceedings of the ACM Web Conference 2025 (WWW '25), April 28--May 2, 2025, Sydney, NSW, Australia}
\acmISBN{979-8-4007-1274-6/25/04}\acmDOI{10.1145/3696410.3714930}

\begin{document}

\title{SCOOT: SLO-Oriented Performance Tuning for LLM Inference Engines}

\author{Ke Cheng}
\authornote{Ke Cheng and Sheng Zhang are with the State Key Laboratory for Novel Software Technology at Nanjing University.}
\authornote{Work done during an internship at Ant Group.}
\authornote{Ke Cheng and Zhi Wang contribute equally to this work.}
\orcid{0000-0003-0336-6916}
\affiliation{
\institution{Nanjing University}
\city{Nanjing}
\country{China}
}
\email{ketonmi@outlook.com}

\author{Zhi Wang}
\authornotemark[3]
\orcid{0000-0002-1095-9632}
\affiliation{
\institution{Ant Group}
\city{Hangzhou}
\country{China}
}
\email{wangchun.wz@antgroup.com}

\author{Wen Hu}
\orcid{0000-0002-4874-2260}
\affiliation{
\institution{Ant Group}
\city{Hangzhou}
\country{China}
}
\email{huwen.hu@antgroup.com}

\author{Tiannuo Yang}
\orcid{0000-0001-5465-9626}
\affiliation{
\institution{Nankai University}
\city{Tianjin}
\country{China}
} 
\email{yangtn@nbjl.nankai.edu.cn}

\author{Jianguo Li}
\orcid{0000-0002-8645-0680}
\authornote{Jianguo Li and Sheng Zhang are corresponding authors.}
\affiliation{
\institution{Ant Group}
\city{Hangzhou}
\country{China}
}
\email{lijg.zero@antgroup.com}

\author{Sheng Zhang}
\authornotemark[1]
\authornotemark[4]
\orcid{0000-0002-6581-6399}
\affiliation{
\institution{Nanjing University}
\city{Nanjing}
\country{China}
} 
\email{sheng@nju.edu.cn}

\renewcommand{\shortauthors}{Ke Cheng et al.}

\begin{abstract}
As large language models (LLMs) are gaining increasing popularity across a wide range of web applications, it is of great importance to optimize service-level objectives (SLOs) for LLM inference services to enhance user satisfaction and improve the competitiveness of cloud vendors. In this paper, we observe that adjusting the parameters of LLM inference engines can improve service performance, and the optimal parameter configurations of different services are different. Therefore, we propose SCOOT, an automatic performance tuning system to optimize SLOs for each LLM inference service by tuning the parameters of the inference engine. SCOOT jointly exploits single-objective and multiple-objective Bayesian optimization (BO) techniques to handle various optimization objectives via exploration and exploitation. Moreover, SCOOT prunes the search space with known constraints and adopts a random forest to learn hidden constraints during the tuning process to mitigate invalid exploration. To improve the tuning efficiency, SCOOT utilizes the parallel suggestion to accelerate the tuning process. Extensive experiments demonstrate that SCOOT considerably outperforms existing tuning techniques in SLO optimization while greatly improving the tuning efficiency. Moreover, SCOOT is universally applicable to various LLM inference engines including vLLM and TensorRT-LLM. Currently, SCOOT has already been implemented in the production environment at Ant Group. 
\end{abstract}



\maketitle

\vspace{-0.25cm}
\section{Introduction}
\label{sec:introduction}
With the impressive capabilities demonstrated by large language models (LLMs) across various web applications \cite{min2023recent}, cloud vendors such as Alibaba Cloud and AWS have started offering services of LLM deployment and inference \cite{alibaba_bailian, aws_sagemaker}. Customers (e.g., developers and providers of web applications) can deploy specified LLMs on their exclusive computing resources. Through the application programming interface (API) provided by the cloud vendor, customers' requests can be continuously served by the deployed LLMs.

These LLM inference services run LLM instances with advanced inference engines such as vLLM \cite{vllm} and TensorRT-LLM \cite{tensorrt_llm}, which are equipped with cutting-edge technologies, such as continuous batching \cite{yu2022orca}, paged attention \cite{kwon2023efficient}, and chunked prefill \cite{agrawal2024taming}. These techniques can accelerate inference and improve throughput, thus delivering a high service performance to customers.

To ensure service performance, customers always agree on service level objectives (SLOs) with cloud vendors. SLOs are defined as a series of performance metric constraints, such as requiring that the 95th percentile latency of requests be less than 1 second. If cloud vendors violate these SLOs, they not only have to compensate customers but also face reputation damage. Therefore, appropriately setting SLOs is critical for cloud vendors \cite{xu2024deploying}. They should optimize SLOs (e.g., guarantee a lower tail latency) to improve customer satisfaction while ensuring that there won’t be SLO violations.

Typically, cloud vendors stress test services with high request rates and set the performance metrics achieved under the stress testing as SLOs. This ensures that SLOs won’t be violated even under heavy workload scenarios. By improving the service performance under stress testing, cloud vendors can deliver better SLOs to customers. 
In this paper, we observe that adjusting the parameters of LLM inference engines has great potential to improve service performance, and the optimal parameter configurations of various LLM inference services are different. Therefore, we stand in the shoes of cloud vendors and optimize SLOs for LLM inference services by tuning inference engine parameters under stress testing.

Numerous performance tuning methods have been proposed and applied across various fields \cite{herodotou2020survey}, but they all fall short of both efficiency and optimality for tuning LLM inference engines. Methods like random sampling and meta-heuristic algorithms, such as Monte Carlo sampling \cite{homem2014monte} and genetic algorithms \cite{katoch2021review} lack efficiency as they fail to fully utilize historical information. Besides, heuristic searches rely on expert knowledge and elaborate pre-profiling to model relationships between parameters and performance. Nevertheless, with the rapid evolution of technologies in LLM inference engines, the parameters' numbers and ranges are frequently updated, which makes the modeled parameter-performance relationship outdated and the designed heuristic search methods inapplicable. Additionally, learning-based approaches such as reinforcement learning (RL) \cite{padakandla2021survey} and Bayesian optimization (BO) \cite{wang2023recent} have also been widely exploited in performance tuning. They can effectively leverage historical information and tune parameters automatically without prior knowledge. However, RL requires a time-consuming training process, while existing BO solutions fail to address the following three challenges.

\textbf{Challenge 1: Various Optimization Objectives.} Customers seek to enhance different performance metrics depending on their application requirements, such as improving request throughput for the periodically invoked offline recommendation application, reducing request tail latency for the online classification application, and minimizing time-to-first-token (TTFT) and time-per-output-token (TPOT) simultaneously for interactive applications such as chatbot, requiring the tuner to have the capability of handling both single-objective and multi-objective optimization problems.

\textbf{Challenge 2: Complex Known and Hidden Constraints.} Some parameters of inference engines depend on the settings of other parameters, thus causing constraints on the search space. For example, for vLLM, {\verb|max-num-batched-tokens|}  must be greater than or equal to {\verb|max-num-seqs|}. We refer to these as known constraints, which can be provided to the tuner ahead of time. Besides, given a specific service, certain parameter combinations can lead to inference engine crashes during stress testing. In this paper, these infeasible parameter combinations are referred to as hidden constraints. Different inference services have different hidden constraints that are initially unknown and must be learned throughout the tuning process. Specifically, for some certain services, vLLM often crashes due to a timeout error during the stress testing when {\verb|scheduler-delay-factor|} is set to a large value.

\textbf{Challenge 3: High Evaluation Overhead.} Learning-based tuners always learn the correlation between the service performance and the engine's parameter settings by evaluating various parameter configurations. As stress testing an inference service takes about 5 to 10 minutes, even with only 30 evaluations for tuning, the total time spent tuning a single service can range from 2.5 to 5 hours. Since the increasing popularity of LLMs leads to the deployment of a large number of LLM inference services, the cumulative time required for tuning these services is prohibitive.

\begin{table}[t]
\centering
\setlength{\abovecaptionskip}{-0.0125cm} 
\caption{vLLM PARAMETERS TO TUNE.}
\begin{tabular}{c|c|c}
	\hline
	Configuration Parameter & Type   	& Range  	                  \\ \hline\hline
	\verb|tensor-parallel|          & Integer			& [1, \#GPUs]       	           	            \\ \hline
	\verb|max-num-seqs|          & Integer			& [64, 8192]       	         	            \\ \hline
	\verb|max-num-batched-tokens|		   & Integer   & [64, 8192]       	      				\\ \hline
	\verb|block-size|          & Enumeration			& \{\verb|8|, \verb|16|, \verb|32|\}	 				\\ \hline
	\verb|scheduler-delay-factor|		   & Float			& [0, 2]			 			\\ \hline
	\verb|enable-chunked-prefill|		   & Boolean			& \{\verb|True|, \verb|False|\}						\\ \hline
	\verb|enable-prefix-caching|		   &  Boolean			& \{\verb|True|, \verb|False|\}			 			\\ \hline
	\verb|disable-custom-all-reduce|		   &  Boolean			& \{\verb|True|, \verb|False|\}			 			\\ \hline
	\verb|use-v2-block-manager|		   &  Boolean			& \{\verb|True|, \verb|False|\}		 			\\ \hline		            
\end{tabular}
\label{tab:tune_param}
\end{table} 

To tackle these challenges, we propose SCOOT, a \underline{\textbf{S}}ervi\underline{\textbf{C}}e-level \underline{\textbf{O}}bjective \underline{\textbf{O}}riented performance \underline{\textbf{T}}uning system, which automatically tune parameters of LLM inference engines to optimize SLOs for LLM inference services. We first propose a general formulation of the inference engine tuning problem to accommodate various optimization objectives and complex constraints, and SCOOT can resolve the problem with BO, where single-objective BO (SOBO) and multi-objective BO (MOBO) \cite{khan2002multi, yang2019multi, daulton2020differentiable} are respectively employed to search optimized parameter configurations for single-objective and multi-objective optimization scenarios, thus addressing challenge 1.
Since constraint violations result in invalid observations caused by engine crashes, which greatly hurts the tuning efficiency, SCOOT prunes the search space with known constraints and exploits a random forest to learn hidden constraints during the tuning process, thus mitigating challenge 2. To resolve challenge 3, SCOOT employs the parallel suggestion technique to recommend multiple parameter configurations each time for simultaneous evaluation, thus fully utilizing idle computing resources to speed up tuning. SCOOT can support various LLM inference engines including vLLM and TensorRT-LLM. Nowadays, SCOOT is in use at Ant Group.

We conduct extensive experiments with various LLMs and different types and numbers of GPUs under request traces collected from various LLM-based web applications at Ant Group. The results show that SCOOT can speed up the tuning process and significantly optimize SLOs, improving the request throughput by up to 68.3\%, reducing the request tail latency by up to  40.6\%, and reducing the TTFT and TPOT by up to 99.8\% and 61.0\%, respectively, compared to the default parameter configuration and existing tuning methods. SCOOT is open-sourced at {\href{https://github.com/Ketonmi/SCOOT}{https://github.com/Ketonmi/SCOOT}. The main contribution of this paper is summarized as follows. 
\begin{itemize}
\item To the best of our knowledge, this is the first study that introduces performance tuning into the field of LLM serving, and we uncover the significance of tuning LLM inference engines with real-world request traces.
\item We propose a general formulation of the inference engine tuning problem that accommodates various optimization objectives and constraints, and we design SCOOT to solve the problem by intelligently searching optimized parameter configurations with BO.
\item Random forest regression is employed by SCOOT to learn hidden constraints during the tuning process to avoid invalid explorations, while the parallel suggestion technique is adopted to significantly improve the tuning efficiency using additional computing resources.
\item Extensive experiments are conducted to confirm the superiority of SCOOT in terms of both the optimality and efficiency for tuning LLM inference engines under various LLMs, computing resources, and request traces.
\end{itemize}

\section{Background and Motivation}
\label{sec:motivation}
\textbf{Background: parameters of the LLM inference engine.}  To provide flexibility of use, inference engines expose many parameters. For vLLM, these parameters include boolean variables such as {\verb|enable-chunked-prefill|} that can enable the chunked prefill technique, integer and float variables such as {\verb|max-num-seqs|} and {\verb|scheduler-delay-factor|} that can change the request scheduling strategy, and enumeration variables such as {\verb|block-size|} that can change the memory allocation policy. In this paper, we focus on tuning parameters that do not affect the accuracy of LLMs. Therefore, we do not consider parameters related to model compression, such as {\verb|quantization|}. Besides, although speculative decoding has been theoretically proven not to hurt LLM accuracy \cite{leviathan2023fast}, it still affects the LLMs' generation results, making it inapplicable for certain applications. Hence, we also do not tune parameters related to speculative decoding. Due to the open-source nature, ease of use, and popularity of vLLM, it is primarily utilized as the inference engine in this paper, and the vLLM parameters to be tuned are listed in Table \ref{tab:tune_param}, which constructs a huge search space of over one hundred billion configuration points.

\begin{table}[t]
\centering
\setlength{\abovecaptionskip}{-0.01cm} 
\caption{VARIOUS SERVICE CHARACTERISTICS.}
\begin{tabular}{c|c|c|c|c}
	\hline
	Service & Application   	& GPU Type  	 & GPU Number   	& LLM                  \\ \hline\hline
	A          & SQL			& A100       	 & 2          	& LLAMA2-13B            \\ \hline
	B		   & \textbf{BOT}   & A100       	 & 2          	& LLAMA2-13B			\\ \hline
	C          & BOT			& \textbf{A10}	 & 2			& LLAMA2-13B			\\ \hline
	D		   & BOT			& A10			 & \textbf{4}	& LLAMA2-13B			\\ \hline
	E		   & BOT			& A10			 & 4			& \textbf{LLAMA2-7B}	\\ \hline		            
\end{tabular}
\vspace{-0.25cm}
\label{tab:param_conf}
\end{table}

\begin{figure}[t]
\centering
\includegraphics[width=0.95\linewidth]{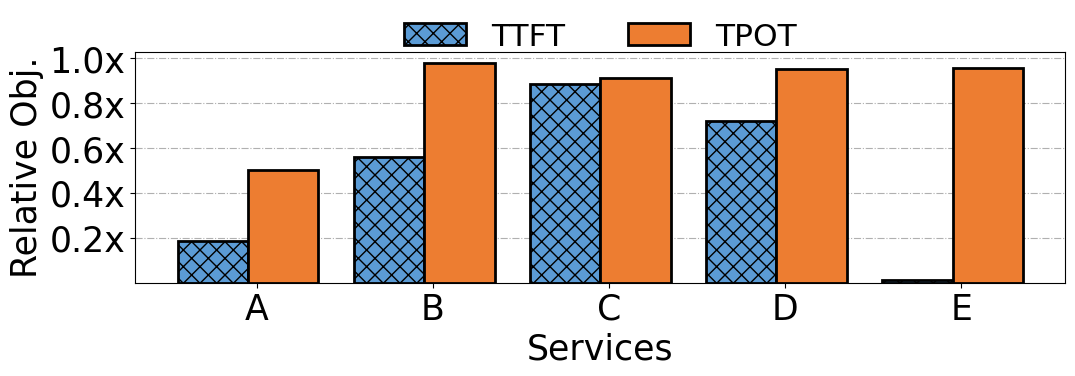}
\vspace{-0.125cm} 
\caption{Optimal TTFT and TPOT for various services. The TTFT and TPOT shown are relative values compared to those of the default parameter configuration. The lower, the better.}
\vspace{0.3cm}
\label{fig:mtv_gain}
\end{figure}

\noindent\textbf{Motivation: parameter adjustment of the LLM inference engine can enhance performance for LLM inference services.}
We conduct experiments to find the optimal parameter configurations of vLLM with grid searches for five services of various LLMs deployed on different numbers and types of GPUs under request traces of various applications. The service characteristics in the experiments are shown in Table \ref{tab:param_conf}, where A10 and A100 separately represent the NVIDIA A10 24G GPU and NVIDIA A100 80G GPU, and SQL and BOT represent the request traces of text-to-SQL and chatbot applications at Ant Group. Other experimental settings are the same as those in Section \ref{sec:experiment}. Figure \ref{fig:mtv_gain} presents the relative TTFT and TPOT performed by optimal parameter configurations for these services. We can observe that TTFT and TPOT can be reduced by up to 98.9\% and 49.9\% compared to the default configuration, respectively, which confirms the significance of performance tuning for LLM inference engines.

\noindent\textbf{Motivation: different inference services' optimal parameter configurations are different.}
We conduct experiments to apply the optimal parameter configuration of a service to other services. Figure \ref{fig:mtv_variance} presents the experimental results. We can observe that the performance exhibited by the optimal parameter configuration for a service may perform poorly or even violate hidden constraints and cause errors for other services. Hence, there is no one ``best practice" configuration that works best in all scenarios, and it is essential to conduct performance tuning for each inference service.

\vspace{-1cm}
\section{Problem Formulation}
\label{sec:formulation}
Given an inference service, the goal of performance tuning for the LLM inference engine is to find the optimal or near-optimal parameter configuration that maximizes the objective and satisfies known and hidden constraints. As we mentioned before, different customers want to optimize different performance metrics. Therefore, we provide a generalized problem formulation that supports various optimization objectives and complex constraints.

\begin{figure}[t!] 
\centering
\begin{subfigure}{0.975\linewidth}
	\centering
	\includegraphics[width=1\linewidth]{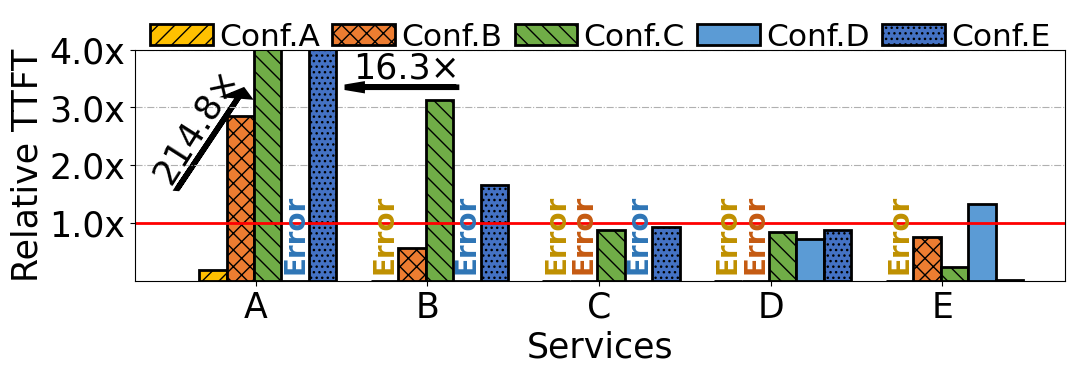}
	\setlength{\abovecaptionskip}{-0.4cm} 
	\setlength{\belowcaptionskip}{-0.01cm} 
	\caption{Relative time-to-first-token (TTFT) compared to the default configuration. The lower, the better.}
	\label{fig:mtv_variance_ttft}
\end{subfigure}
\centering
\begin{subfigure}{0.975\linewidth}
	\centering
	\includegraphics[width=1\linewidth]{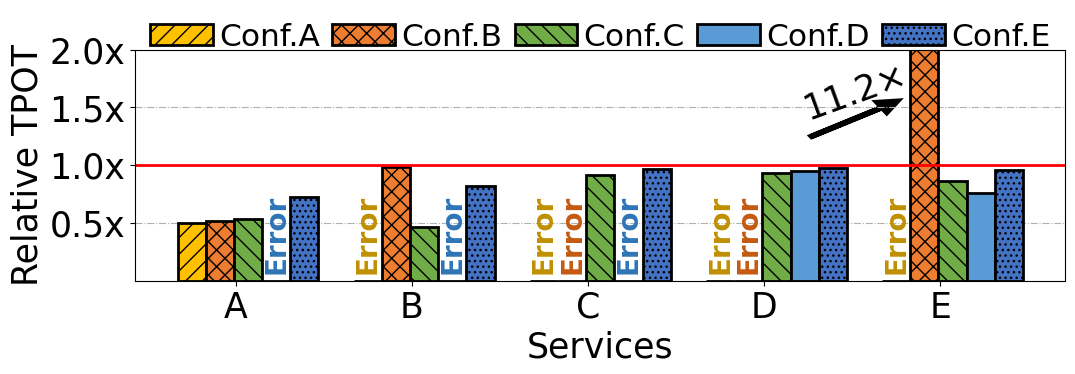}
	\setlength{\abovecaptionskip}{-0.4cm} 
	\caption{Relative time-per-output-token (TPOT) compared to the default configuration. The lower, the better.}
	\label{fig:mtv_variance_tpot}
\end{subfigure}
\setlength{\abovecaptionskip}{-0.0005cm} 
\caption{TTFT and TPOT for applying optimal parameter configurations of different services to other services. Conf. $X$ represents the optimal parameter configuration for the service $X\in\{A,B,C,D,E\}$. The red lines (1.0×) in Sub-figures (a) and (b) indicate the TTFT and TPOT under the default parameter configuration of the inference engine, respectively.}
\vspace{0.25cm}
\label{fig:mtv_variance}
\end{figure}

\begin{figure*}[t]
\centering
\includegraphics[width=1\linewidth]{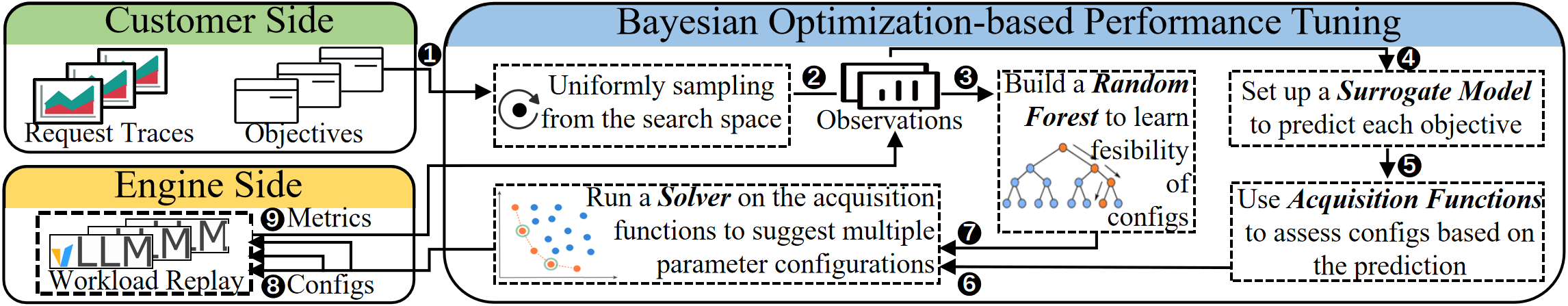}
\setlength{\abovecaptionskip}{-0.25cm} 
\caption{SCOOT workflow. SCOOT leverages BO to find optimized parameter configurations via exploration and exploitation.}
\vspace{-0.25cm}
\label{fig:scoot_workflow}
\end{figure*}

We define the configuration search space $\boldsymbol{\Lambda}=\Lambda_1\times \Lambda_2\times...\times \Lambda_n$, where $\Lambda_i$ represent the range of the $i$th parameter to tune. Vector $\boldsymbol{x}\in \boldsymbol{\Lambda}$ is utilized to denote a specific parameter configuration, where the $i$th element $\boldsymbol{x}[i]\in \Lambda_i$ corresponds to the value of the $i$th parameter. Moreover, we respectively use $T(\boldsymbol{x})$, $L(\boldsymbol{x})$, $\Phi(\boldsymbol{x})$, and $\Theta(\boldsymbol{x})$ to represent the functions of request throughput, tail latency, average TTFT, and average TPOT for the inference service under configuration $\boldsymbol{x}$. We leverage $\mathcal{C}=\{c_1,c_2,...,c_m\}$ to symbolize the set of known constraints, where $c_i$ is the $i$th known constraint which restricts the relationship between elements of $\boldsymbol{x}$. For example, a known constraint can be expressed as  ``$\boldsymbol{x}[1]<\boldsymbol{x}[3]\ \text{if}\ \boldsymbol{x}[4]\ \text{is}\  \text{False}$". Moreover, we utilize $POF(\boldsymbol{x})$ to represent $\boldsymbol{x}$'s probability of feasibility (POF) that $\boldsymbol{x}$ won't violate hidden constraints. Therefore, the tuning problem is formulated as follows:

\begin{equation}
\mathbb{P}: \max\limits_{\boldsymbol{x}\in \boldsymbol{\Lambda}}\ \ \lambda_{t}\cdot T(\boldsymbol{x}), \lambda_{l}\cdot L(\boldsymbol{x}), \lambda_{\phi}\cdot \Phi(\boldsymbol{x}), \lambda_{\theta}\cdot \Theta(\boldsymbol{x})
\label{eq:tuning_formulation}
\end{equation}
\textbf{s. t.}
\begin{equation}
c_i, \forall c_i\in \mathcal{C}, 
\label{eq:known_constr}
\end{equation}
\begin{equation}
POF(\boldsymbol{x})\ge \Delta, 
\label{eq:hidden_constr}
\end{equation}
where Eq. (\ref{eq:known_constr}) and Eq. (\ref{eq:hidden_constr}) respectively denotes the known and hidden constraints. $\Delta$ is the POF threshold used to avoid violations of hidden constraints. $\lambda_t\in \{0, 1\}$ and $\lambda_{l}, \lambda_\phi, \lambda{\theta} \in \{0, -1\}$ are utilized to control the optimization objectives. If $(\lambda_t, \lambda_l, \lambda_\phi, \lambda_\theta)=(1, 0, 0, 0)$, the optimization objective is to only maximize the request throughput, which is applicable to periodically invoked applications such as offline recommendation. If $(\lambda_t, \lambda_l, \lambda_\phi, \lambda_\theta)=(0, -1, 0, 0)$, the optimization objective is to only minimize the tail latency, which is appropriate for non-interactive online applications such as classification. If $(\lambda_t, \lambda_l, \lambda_\phi, \lambda_\theta)=(0, 0, -1, -1)$, the optimization objective is to simultaneously minimize the TTFT and TPOT, which is suitable for interactive applications.

\section{SCOOT: Solution Description}
\label{sec:solution}
We present SCOOT’s design, including its use of SOBO and MOBO for solving $\mathbb{P}$, its innovative features of parallel suggestion and random forest-based POF learning, and the SLO robustness assurance.

\subsection{SCOOT Workflow}
The workflow of SCOOT is depicted in Fig. \ref{fig:scoot_workflow}, which consists of nine key steps to find optimized parameter configurations. Customers define their optimization objectives, and SCOOT \ding{182} gathers their request traces that include both input text of requests and output text generated by the LLM. 
Then, SCOOT leverages Sobol sequence-based Quasi-Monte Carlo \cite{caflisch1998monte} to uniformly \ding{183} sample configurations across the search space, where the number of samples matches the search space’s dimensionality. Then, SCOOT runs the inference engine with each sampled configuration and stress tests the inference service to obtain initial observations.
These observations of configuration-performance pairs are leveraged to \ding{184} build a random forest and \ding{185} construct a surrogate model to learn the $POF(\cdot)$ and predict the probability distribution of each optimization objective, respectively. Subsequently, acquisition functions are exploited to \ding{186} assess configurations according to the predicted results. Based on the assessment, a solver \ding{187} suggests multiple configurations in parallel while adhering to the known constraints and \ding{188} ensuring hidden constraints using $POF(\cdot)$ learned by the random forest. Lastly, the inference engine \ding{189} is started with each suggested parameter configuration, and the stress testing is conducted to \ding{190} obtain new observations to refine the random forest and the surrogate model. Steps \ding{184}$\sim$\ding{190} run iteratively, and the tuning process stops until the number of observations reaches a given threshold. 

In the workflow, the LLM, GPU number, and GPU type used for tuning are the same as the inference service owned by the customer.

\vspace{-0.05cm}
\subsection{Bayesian Optimization-based Solution}
BO is a theoretically grounded method for finding the optimum of black-box functions. It can explore the complex multi-dimensional search space efficiently and intelligently \cite{wang2023recent}, which is suitable for solving problems with expensive evaluation overhead. BO leverages a surrogate model to approximate the objective function and use acquisition functions to assess configuration points for suggesting.  

SCOOT leverages SOBO to maximize throughput and minimize tail latency for non-interactive offline and online applications, respectively. For interactive applications, it exploits MOBO to minimize TTFT and TPOT simultaneously by finding a set of parameter configurations representing the Pareto frontier that denotes the optimal trade-offs between TTFT and TPOT. We do not linearly combine TTFT and TPOT as a single optimization objective and solve it with SOBO because TTFT can be hundreds or even thousands of times larger than TPOT. Hence, it is difficult to assign weights to TTFT and TPOT to make a good trade-off. Besides, TTFT and TPOT are always two conflicting optimization objectives as illustrated in Appendix \ref{sec:conflict}.

\subsection{Surrogate Model}
The surrogate model of BO predicts the objective function $f(\boldsymbol{x})$ based on observations. It models $f(\boldsymbol{x})$ for a given configuration $\boldsymbol{x}$ as a random variable and predicts its probability distribution. In the context of LLM inference engine tuning, $f(\boldsymbol{x})$ can represent the objective functions of request throughput $T(\boldsymbol{x})$, request tail latency $L(\boldsymbol{x})$, average TTFT $\Phi(\boldsymbol{x})$, and average TPOT $\Theta(\boldsymbol{x})$.

SCOOT uses the Gaussian process (GP) as the surrogate model. Given a parameter configuration $\boldsymbol{x}$, GP assumes that the probability distribution of $f(\boldsymbol{x})$ follows a Gaussian distribution whose mean $\mu(f(\boldsymbol{x}))$ and the variance $\sigma^2(f(\boldsymbol{x}))$ are respectively computed by
\begin{equation}
\mu(f(\boldsymbol{x}))=k(\boldsymbol{x}, \boldsymbol{\mathcal{X}}) (k(\boldsymbol{\mathcal{X}}, \boldsymbol{\mathcal{X}}) + \tau^2 I)^{-1} \boldsymbol{\mathcal{Y}},
\end{equation}
\vspace{-0.5cm}
\begin{equation}
\sigma^2(f(\boldsymbol{x}))=k(\boldsymbol{x}, \boldsymbol{x}) - k(\boldsymbol{x}, \boldsymbol{\mathcal{X}})(k(\boldsymbol{\mathcal{X}}, \boldsymbol{\mathcal{X}}) + \tau^2 I)^{-1} k(\boldsymbol{\mathcal{X}}, \boldsymbol{x}),	
\end{equation}
where $\boldsymbol{\mathcal{X}}$ represents the previously evaluated configurations, $\boldsymbol{\mathcal{Y}}$ denotes the corresponding observed objective values, $\tau^2$ is the level of white noise, and $k(\boldsymbol{x}, \boldsymbol{x'})$ is the covariance function that quantifies the similarity between input points $\boldsymbol{x}$ and $\boldsymbol{x'}$ for inferring the relationships between their objective function values. SCOOT employs the Matern kernel ($\frac{3}{2}$) with input wrapping \cite{snoek2014input} as the covariance function due to its capability to balance the smoothness and flexibility when modeling unknown functions. SCOOT utilizes maximum likelihood estimation \cite{seeger2004gaussian} to learn $\tau^2$ during tuning.

For SOBO, a single-output GP is leveraged to predict the objective function. For MOBO, SCOOT adopts a multi-output GP by considering each output to be independent. During the tuning process, the prediction accuracy of the GP model is continuously improved as more observations are collected.

\vspace{-0.05cm}
\subsection{Acquisition Function}
\vspace{-0.005cm}
The acquisition function of BO assesses parameter configurations in the search space, which calculates a score for each configuration point $\boldsymbol{x}$ according to the surrogate model's predicted mean $\mu(f(\boldsymbol{x}))$ (indicating the expected performance) and variance $\sigma^2(f(\boldsymbol{x}))$ (representing uncertainty). Parameter configurations with high scores are more likely to be suggested for evaluation. Different acquisition functions balance exploration (visiting areas with high uncertainty) and exploitation (intensively searching areas with good known objective values) in different manners.

\vspace{-0.005cm}
\subsubsection{\textbf{SOBO Acquisition Function}}
For SOBO, commonly used acquisition functions include upper confidence bound (UCB), probability of improvement (PI), and expected improvement (EI). UCB incorporates both mean and variance into the score and prioritizes the point with a high balance of exploration and exploitation through a trade-off parameter $\beta$, which is expressed by
\begin{equation}
UCB(\boldsymbol{x})=\mu(f(\boldsymbol{x})) + \beta\cdot \sigma(f(\boldsymbol{x})),
\label{eq:ucb}
\end{equation} where $\sigma(f(\boldsymbol{x}))=\sqrt{\sigma^2(f(\boldsymbol{x}))}$ represents the standard deviation.

PI prioritizes the point that is likely to yield an improvement over the current known best observation, which is expressed by 
\begin{equation}
PI(\boldsymbol{x})=P(f(\boldsymbol{x})+\xi > f(\boldsymbol{x}^+)),
\label{eq:pi}
\end{equation}
where $\boldsymbol{x}^{+}$ is the point that has the largest objective function value known so far, and $\xi$  is leveraged to encourage exploration.

EI not only considers the probability of objective improvement but also the magnitude of the expected improvement, which is computed by
\begin{equation}
EI(\boldsymbol{x})=\mathbb{E}(\max(0, f(\boldsymbol{x}) +\xi - f(\boldsymbol{x}^{+}))),
\label{eq:ei}
\end{equation}
where $\boldsymbol{x}^{+}$ represents the best observed configuration point as well, and $\xi$ is also a parameter used to encourage exploration. 

\begin{figure}[t]
\centering
\begin{subfigure}{0.35\linewidth}
	\centering
	\includegraphics[width=1\linewidth]{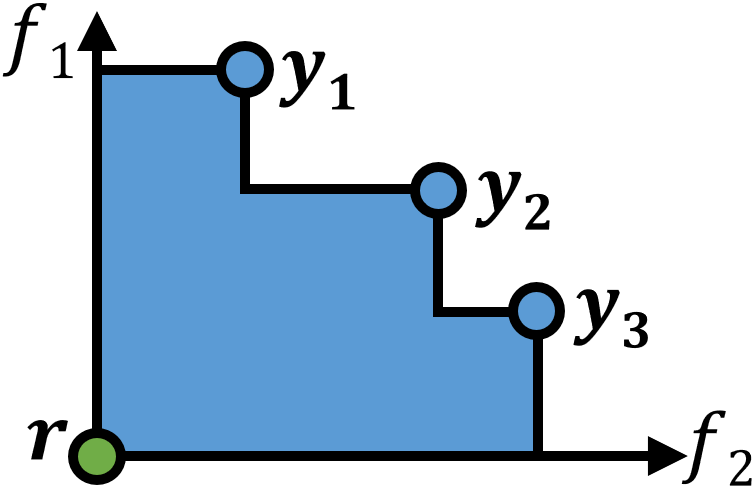}
	\caption{Hypervolume (HV).}
	\label{fig:hv}
\end{subfigure}
\centering
\begin{subfigure}{0.35\linewidth}
	\centering
	\includegraphics[width=1\linewidth]{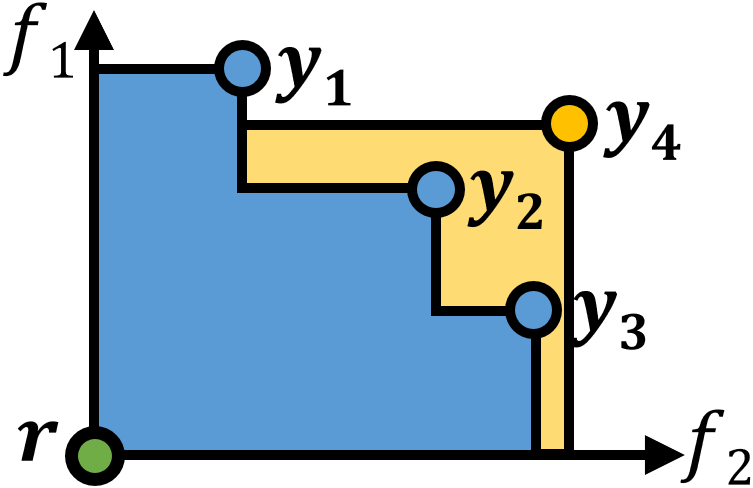}
	\caption{HV improvement.}
	\label{fig:hvi}
\end{subfigure}
\setlength{\abovecaptionskip}{-0.0125cm} 
\caption{Illustration of two-dimensional HV. $f_1$ and $f_2$ are two objective functions and $\boldsymbol{r}$ is the reference point. (a) Blue area represents the HV of the existing solution set. (b) Yellow area depicts the HV improvement after adding $\boldsymbol{y}_4$.}
\vspace{0.25cm}
\label{fig:hv_example}
\end{figure}

\subsubsection{\textbf{MOBO Acquisition Function}}
For MOBO, expected hypervolume improvement (EHVI) is widely adopted as the acquisition function to identify promising parameter configurations that denote optimal trade-offs between multiple optimization objectives. 
EHVI assesses a new configuration point based on the potential improvement it might bring to the hypervolume of the existing solution set, which is calculated by 
\begin{equation}
EHVI(\boldsymbol{x})=\mathbb{E}(\max(0,HV(\boldsymbol{\mathcal{Y}}\cup\{f(\boldsymbol{x})\})-HV(\boldsymbol{\mathcal{Y}}))),
\label{eq:ehvi}
\end{equation}
where $HV(\cdot)$ is the hypervolume function and $\boldsymbol{\mathcal{Y}}$ denotes the objective values of previously evaluated configurations. 

The hypervolume function $HV(\cdot)$ is calculated by measuring the volume of space enclosed between the Pareto frontier and a reference point $\boldsymbol{r}$, where $\boldsymbol{r}$ is typically a lower bound point for all objective functions. When there are only two optimization objectives, $HV(\cdot)$ is calculated by summing the areas of rectangles formed between each point on the Pareto frontier and $\boldsymbol{r}$. Figure \ref{fig:hv_example} presents how the two-dimensional $HV(\cdot)$ is computed.

A larger $HV(\cdot)$ value indicates a Pareto front that covers a larger objective space, offering a broader range of promising solutions. By maximizing the EVHI, the search process can be guided towards configurations that can improve the overall quality and diversity of the Pareto frontier, thus optimizing multiple objectives simultaneously and providing more optional configurations for customers.

\vspace{-0.01cm}
\subsection{Configuration Suggestion}
\subsubsection{\textbf{SOBO Suggestion}} For SOBO, a common practice is to select a specific acquisition function and suggest parameter configurations that maximize the acquisition function. However, when it comes to real-world applications, the selected acquisition function might work sub-optimally for the tuning task, and the best acquisition function is challenging to identify beforehand \cite{cowen2022hebo}. 

To mitigate this issue, SCOOT employs multi-objective acquisition function ensemble (MACE) \cite{zhang2021efficient}. MACE uses three acquisition functions, UCB, PI, and EI, to assess configuration points and runs a solver to find out the Pareto frontier that denotes the optimal trade-offs of the three acquisition functions. Suggested configurations are randomly selected in the Pareto frontier. With MACE, SCOOT can avoid using sub-optimal acquisition functions all the time, thereby improving the quality of suggested parameter configurations. 

During the tuning process, for UCB, $\beta$ is dynamically adjusted, and in the $t$th tuning iteration, $\beta$ is calculated by ${2log(\frac{t^2\pi^2}{6\delta})}$, where $\delta=\frac{1}{t^2}$ \cite{srinivas2010gaussian}. For EI and PI, $\xi$ is set to a fixed value of 0.0001.

\vspace{-0.01cm}
\subsubsection{\textbf{MOBO Suggestion}} For MOBO, SCOOT leverages the EHVI as the acquisition function and runs a solver to suggest parameter configurations by maximizing EHVI. The default configuration of the inference engine is chosen as the reference point $\boldsymbol{r}$.

\vspace{-0.01cm}
\subsubsection{\textbf{Known Constraint Assurance}}
For both SOBO and MOBO, in the process of solvers solving optimization problems, the solution space is pruned by known constraints.

\vspace{-0.01cm}
\subsection{Parallel Configuration Suggestion}
With the popularity of LLMs, an increasing number of LLM inference services are being deployed, and tuning them incurs substantial time overhead. To mitigate this problem, SCOOT suggests multiple parameter configurations at a time and utilizes multiple sets of computing resources to conduct stress testing in parallel. 

Given the parallelism degree $k$, for SOBO, SCOOT randomly selects $k$ points from the Pareto frontier of the three acquisition functions for suggesting. For MOBO, the top-$k$ points with the largest EHVI are suggested. Thus, tuning can be accelerated by a factor of $k$ when the total observation number is fixed.

Although the parallel suggestion requires additional computing resources to evaluate configurations in parallel, many computing resources in production clusters are often idle during off-peak hours such as nights and weekends, which can be conveniently used for tuning LLM inference engines.

\begin{algorithm}[t]
\caption{SCOOT: BO-based Performance Tuning}
\label{alg:threshold_adjust}
\KwIn{$N$: total observation number;
	$k$: parallelism degree;}
\DontPrintSemicolon

\tcp{\textcolor{blue}{Initialize observations}}
$\mathcal{O} \gets$ UNIFORM\_SAMPLE\_AND\_EVALUATE()

$RF\gets$ TRAIN($\mathcal{O}$) \tcp{\textcolor{blue}{Train random forest regression}}

$\Delta \gets 0.5$; $c\gets0$; $\upsilon\gets0.05$

\While{$|\mathcal{O}|< N$}{
	\tcp{\textcolor{blue}{Suggest $k$ configurations}}
	$\{\boldsymbol{x_1},...,\boldsymbol{x_k}\}\gets$ SUGGEST($k$,  $\mathcal{O}$, $RF$, $\Delta$)
	
	\tcp{\textcolor{blue}{Collect performance metrics}}
	$\{\boldsymbol{y_1},...,\boldsymbol{y_k}\}\gets$ EVALUATE($\{\boldsymbol{x_1},...,\boldsymbol{x_k}\}$) 
	
	\tcp{\textcolor{blue}{Adjust feasibility threshold}}
	\uIf{\text{there} are invalid $\boldsymbol{y}$ in $\{\boldsymbol{y_1}, ..., \boldsymbol{y_k}\}$}{
		$\Delta\gets\min(0.75, \max(0.5, \Delta+\upsilon))$
		
		$c$ $\gets0$	\tcp{\textcolor{blue}{Clear feasible suggestion count}}
	}
	\Else{
		$c \gets c+ k$
		
		\If{$c\ge5$}{
			$\Delta\gets\max(0.25, \Delta-\upsilon)$
			
			$c \gets c-5$
		}
	}
	
	\tcp{\textcolor{blue}{Update observations}}
	$\mathcal{O}\gets$ UNION($\mathcal{O}$, $\{\boldsymbol{x_1},...,\boldsymbol{x_k}\}$, $\{\boldsymbol{y_1},...,\boldsymbol{y_k}\}$) 
	
	$RF\gets$ TRAIN($\mathcal{O}$) \tcp{\textcolor{blue}{Refine Random Forest}}
}
\end{algorithm}

\vspace{-0.25cm}
\subsection{Random Forest-based POF Learning}
\vspace{-0.05cm}
To handle hidden constraints, a common practice is to assign penalty objective values to the infeasible parameter configurations that cause engine crashes. However, it is challenging to set appropriate penalty objective values. Besides, penalty objective values often hurt the surrogate model \cite{hellsten2023baco}. Therefore, SCOOT leverages random forest regression to learn the $POF(\cdot)$ and handle hidden constraints by restricting the solver to adhere to Eq. (\ref{eq:hidden_constr}) in the process of configuration suggestion. In this way, SCOOT can substantially reduce invalid observations without affecting the surrogate model.

For the hidden constraint expressed in Eq. (\ref{eq:hidden_constr}), the feasibility probability threshold $\Delta$ is critical. If $\Delta$ is fixed, setting it too high may result in repeatedly suggesting configuration points near already observed feasible configurations, potentially missing out on other superior configuration points. Conversely, setting $\Delta$ too low may lead to suggestions of infeasible configurations. Hence, SCOOT dynamically adjusts the threshold $\Delta$ during the tuning process. 

The dynamic adjustment process of $\Delta$ can be found in Algorithm \ref{alg:threshold_adjust}, where we can observe that when infeasible parameter configurations are suggested, $\Delta$ is increased to reduce the likelihood of further suggesting infeasible configurations. Besides, after continuously suggesting feasible configurations five times, $\Delta$ is decreased to allow the discovery of potential superior configurations. Additionally, during the tuning process, the random forest is continuously refined using the latest observations to enhance prediction accuracy, ensuring that hidden constraints are intelligently followed without compromising the quality of performance tuning.

Since known and hidden constraints pure the solution space, solvers may be unable to give a sufficient number of solutions during the configuration suggestion. In such scenarios, SCOOT randomly samples configuration points with Sobol sequence-based Quasi-Monte Carlo to make up for the shortfall.

\begin{figure}[t]
\centering
\begin{subfigure}{0.475\linewidth}
	\centering
	\includegraphics[width=1\linewidth]{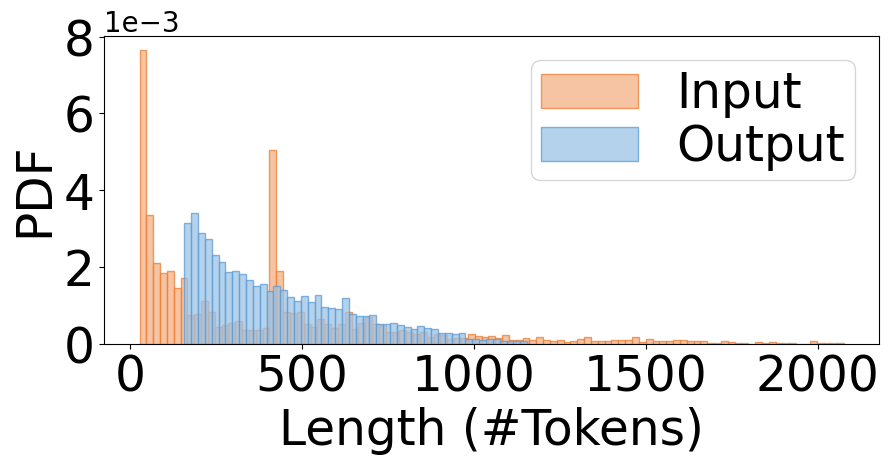}
	\setlength{\abovecaptionskip}{-0.35cm} 
	\setlength{\belowcaptionskip}{-0.1cm} 
	\caption{BOT.}
	\label{fig:bot}
\end{subfigure}
\centering
\begin{subfigure}{0.475\linewidth}
	\centering
	\includegraphics[width=1\linewidth]{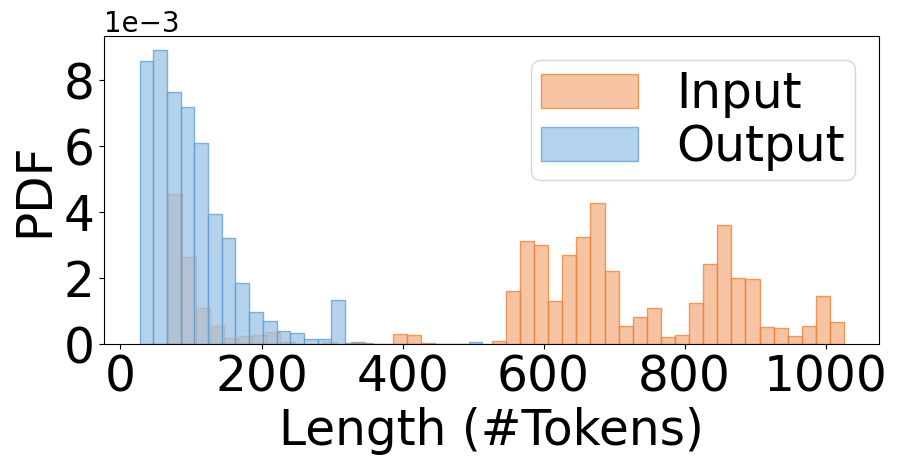}
	\setlength{\abovecaptionskip}{-0.35cm} 
	\setlength{\belowcaptionskip}{-0.1cm} 
	\caption{SQL.}
	\label{fig:sql}
\end{subfigure}
\begin{subfigure}{0.46\linewidth}
	\centering
	\includegraphics[width=1\linewidth]{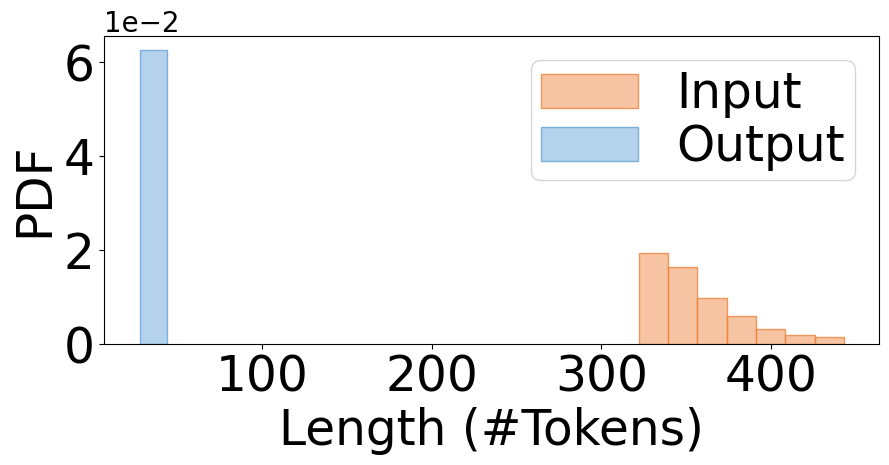}
	\setlength{\abovecaptionskip}{-0.35cm} 
	\caption{CLS.}
	\label{fig:cls}
\end{subfigure}
\centering
\begin{subfigure}{0.485\linewidth}
	\centering
	\includegraphics[width=1\linewidth]{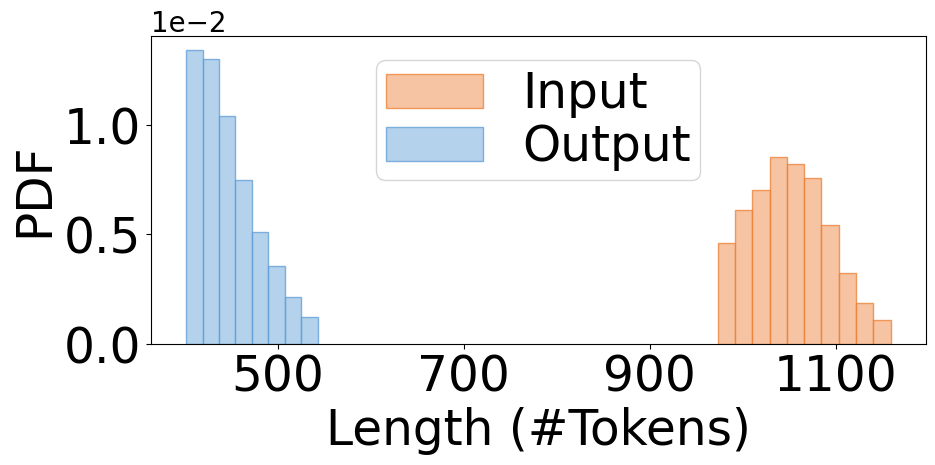}
	\setlength{\abovecaptionskip}{-0.35cm} 
	\caption{REC.}
	\label{fig:rec}
\end{subfigure}
\setlength{\abovecaptionskip}{-0.0125cm} 
\caption{Probability density function (PDF) of the request input and output lengths for four application request traces.}
\label{fig:input_output}
\end{figure}

\vspace{-0.25cm}
\subsection{Ensuring SLO Robustness}
Through extensive experiments, we find that when the performance tuning is conducted under different request arrival orders, the best parameter configurations are almost the same, but the corresponding optimized SLOs vary. Thus, after performance tuning, we conduct stress testing multiple times again using the best configuration, varying request arrival orders, and select the worst-case objectives as the SLOs to ensure SLO Robustness. Finally, we summarize the overall tuning procedure of SCOOT in Algorithm \ref{alg:threshold_adjust}.

\vspace{-0.25cm}
\section{Experimental Evaluation}
\label{sec:experiment}
\subsection{Experiment Setup}
We implement SCOOT on top of HEBO \cite{cowen2022hebo}, where the random forest regression is implemented using the sklearn \cite{pedregosa2011scikit} library. Request traces are collected from four LLM inference services at Ant Group, including applications of text-to-SQL (\textbf{SQL}), chatbot (\textbf{BOT}), classification (\textbf{CLS}), and recommendation (\textbf{REC}). Since the computational load of an LLM inference request depends on its input and output lengths, to enhance the robustness of optimized SLOs, SCOOT uses requests with the longest 50\% of output lengths for stress testing. The input and output lengths of these requests are presented in Fig. \ref{fig:input_output}. 
For SQL and BOT, TTFT and TPOT are optimization objectives at the same time. For CLS and REC, 95th percentile latency and request throughput are utilized as optimization objectives, respectively.

Multiple sets of computing resources are utilized in experiments, including \textbf{2A10}, \textbf{4A10}, \textbf{2A100}, and \textbf{4A100}, where A10 represents the NVIDIA A10 24GB GPU, A100 denotes the NVIDIA A100 80GB GPU, and the integer before the GPU type indicate the number of GPUs. For each set of computing resources, 256GB CPU memory and one 2.90GHz 32-core Intel(R) Xeon(R) CPU are equipped. Besides, A100 GPUs are connected over NVLink while A10 GPUs are connected over PCIE. We use LLAMA2-7B and LLAMA2-13B \cite{llama2} in experiments because the model architecture of LLAMA is representative and fine-tuned 7B and 13B LLMs can fulfill the requirements for most applications in practice. Due to the high usability and popularity of vLLM, we adopt vLLM-0.4.2 as the inference engine to tune and provide its known constraints in Appendix \ref{sec:known_constraints}.

In each configuration evaluation, when the inference engine is started with the suggested configuration, we send requests from the request trace for 100 seconds and the request arrival times are generated using Poisson distribution with various request rates. For A10, the request rates for SQL, BOT, CLS, and REC are 5,5, 10, and 15, respectively. For A100, the request rates of these request traces are twice that of A10. The request rate is set according to the actual workload of the applications as well as considering the computing capability and GPU memory capacity of GPUs.

We compare SCOOT with three baselines, including random sampling (\textbf{RD}), genetic algorithm (\textbf{GA}), and Vanilla BO (\textbf{VBO}). Detailed baseline description is provided in Appendix \ref{sec:baseline}.

\vspace{-0.05cm}
\subsection{SLO Optimization}
In the experiments of SLO optimization, we limit the total observation number to 30 and set the suggestion parallelism degree (PD) of SCOOT to one. The experimental results are presented in Figures \ref{fig:bot_llama7b}, \ref{fig:bot_llama13b}, \ref{fig:sql_llama7b}, \ref{fig:sql_llama13b}, \ref{fig:cls_latency}, and \ref{fig:rec_tp}, where we not only present the SLO improvement under each set of computing resources but also the average SLO improvement across various computing resources. The bar represents the SLO improvement compared to the default parameter configuration, and \textbf{the higher the bars, the better}. In addition,  \textbf{×} represents that no better configuration than the default configuration is found.

For BOT and SQL applications, TTFT and TPOT are optimized simultaneously, and a Pareto frontier is resolved. We choose SLOs of the configuration with maximal HV from the Pareto frontier as the tuning result to report since it represents the best trade-off between TTFT and TPOT. HV of a single configuration point $\boldsymbol{x}$ is calculated by 
\begin{equation}
HV(\boldsymbol{x})=\max(0, y_{\boldsymbol{x}}^{\phi}-y_{\boldsymbol{r}}^{\phi}) \cdot \max(0, y_{\boldsymbol{x}}^{\theta}-y_{\boldsymbol{r}}^{\theta}),
\label{eq:hv_single_point}
\end{equation}
where $y_{\boldsymbol{x}}^{\phi}$, $y_{\boldsymbol{x}}^{\theta}$, $y_{\boldsymbol{r}}^{\phi}$, and $y_{\boldsymbol{r}}^{\theta}$ denote the observed TTFT and TPOT of $\boldsymbol{x}$ and the reference point $\boldsymbol{r}$, respectively.

\begin{figure}[t]
\centering
\begin{subfigure}{0.475\linewidth}
	\centering
	\includegraphics[width=1\linewidth]{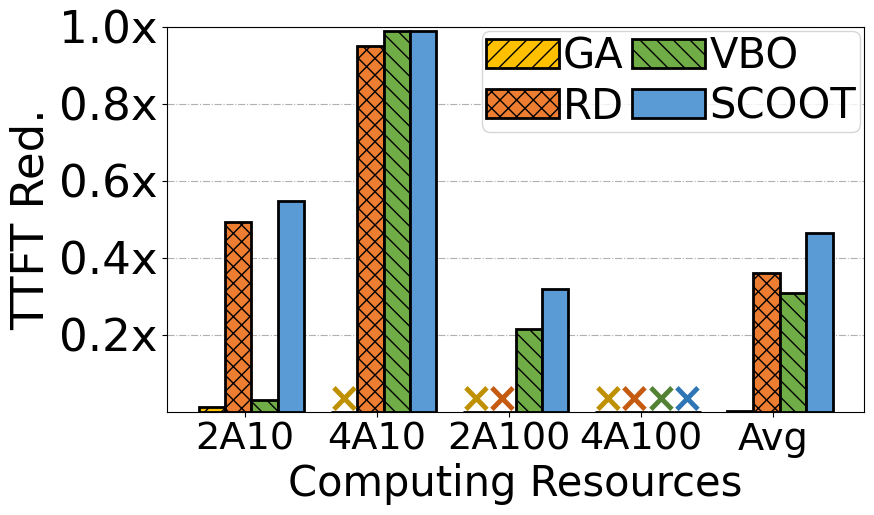}
	\setlength{\abovecaptionskip}{-0.35cm} 
	\caption{TTFT reduction.}
	\label{fig:bot_llama7b_ttft}
\end{subfigure}
\centering
\begin{subfigure}{0.475\linewidth}
	\centering
	\includegraphics[width=1\linewidth]{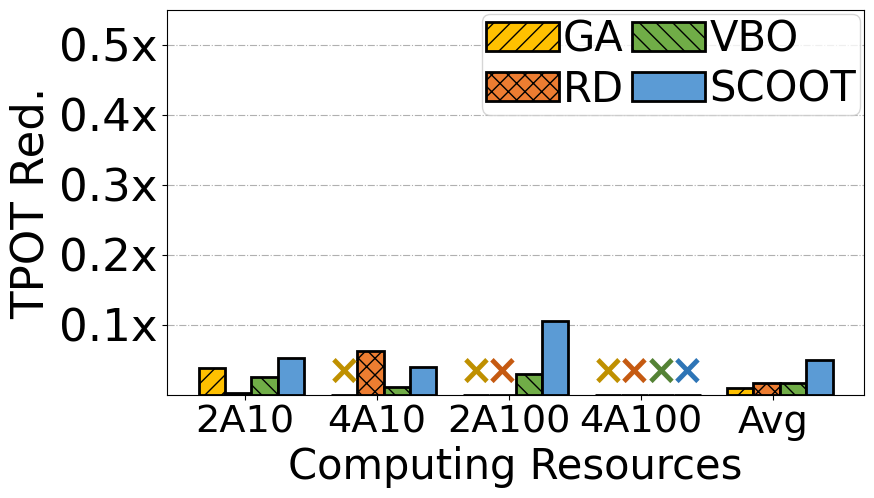}
	\setlength{\abovecaptionskip}{-0.35cm} 
	\caption{TPOT reduction.}
	\label{fig:bot_llama7b_tpot}
\end{subfigure}
\setlength{\abovecaptionskip}{-0.0125cm} 
\caption{SLO optimization for \textbf{BOT} under \textbf{LLAMA2-7B}.}
\label{fig:bot_llama7b}
\vspace{-0.5cm}
\end{figure}

\begin{figure}[t]
\centering
\begin{subfigure}{0.475\linewidth}
	\centering
	\includegraphics[width=1\linewidth]{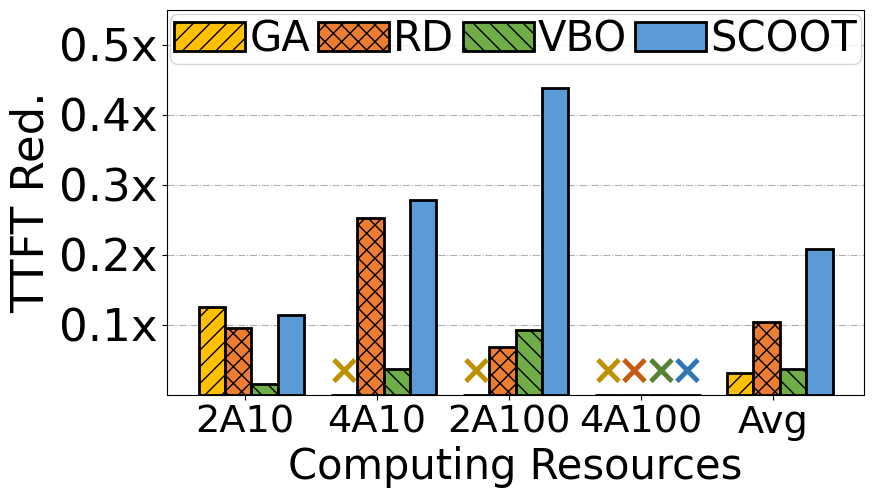}
	\setlength{\abovecaptionskip}{-0.35cm} 
	\caption{TTFT reduction.}
	\label{fig:bot_llama13b_ttft}
\end{subfigure}
\centering
\begin{subfigure}{0.475\linewidth}
	\centering
	\includegraphics[width=1\linewidth]{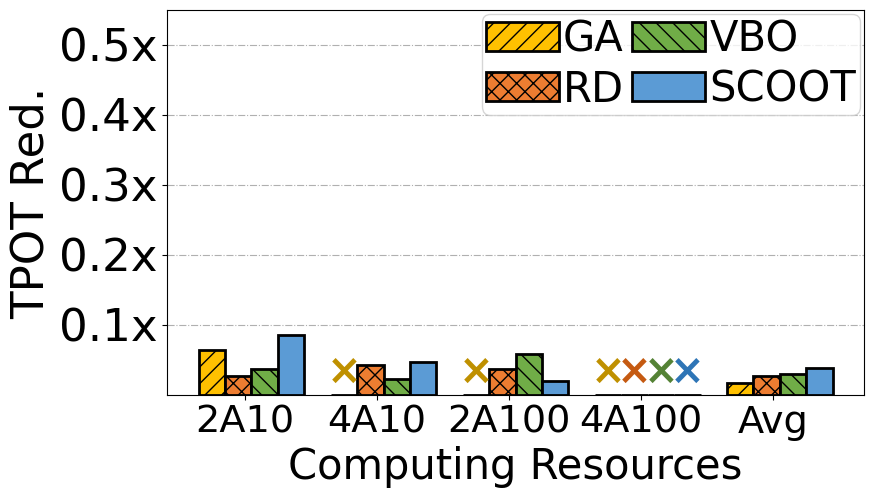}
	\setlength{\abovecaptionskip}{-0.35cm} 
	\caption{TPOT reduction.}
	\label{fig:bot_llama13b_tpot}
\end{subfigure}
\setlength{\abovecaptionskip}{-0.0125cm} 
\caption{SLO optimization for \textbf{BOT} under \textbf{LLAMA2-13B}.}
\label{fig:bot_llama13b}
\vspace{-0.1cm}
\end{figure}

\begin{figure}[t]
\centering
\begin{subfigure}{0.475\linewidth}
	\centering
	\includegraphics[width=1\linewidth]{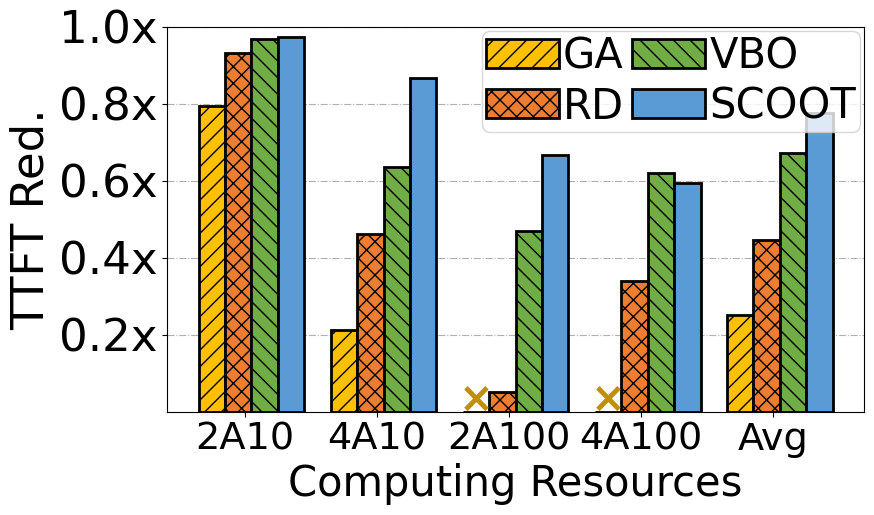}
	\setlength{\abovecaptionskip}{-0.35cm} 
	\caption{TTFT reduction.}
	\label{fig:sql_llama7b_ttft}
\end{subfigure}
\centering
\begin{subfigure}{0.475\linewidth}
	\centering
	\includegraphics[width=1\linewidth]{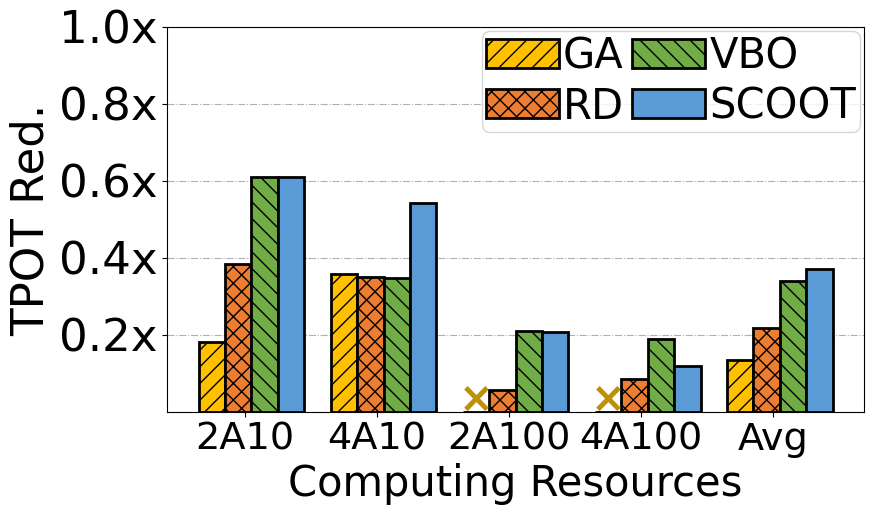}
	\setlength{\abovecaptionskip}{-0.35cm} 
	\caption{TPOT reduction.}
	\label{fig:sql_llama7b_tpot}
\end{subfigure}
\setlength{\abovecaptionskip}{-0.0125cm} 
\caption{SLO optimization for \textbf{SQL} under \textbf{LLAMA2-7B}.}
\label{fig:sql_llama7b}
\vspace{-0.5cm}
\end{figure}

\begin{figure}[t]
\centering
\begin{subfigure}{0.475\linewidth}
	\centering
	\includegraphics[width=1\linewidth]{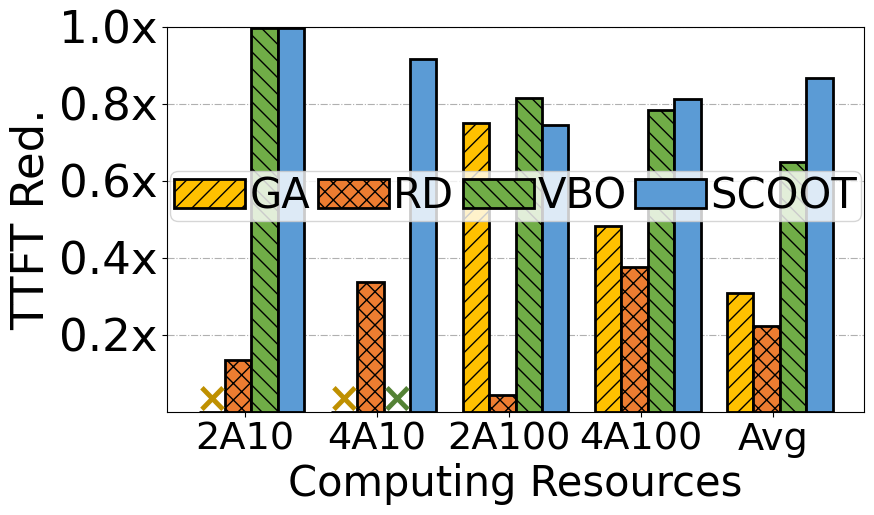}
	\setlength{\abovecaptionskip}{-0.35cm} 
	\caption{TTFT reduction.}
	\label{fig:sql_llama13b_ttft}
\end{subfigure}
\centering
\begin{subfigure}{0.475\linewidth}
	\centering
	\includegraphics[width=1\linewidth]{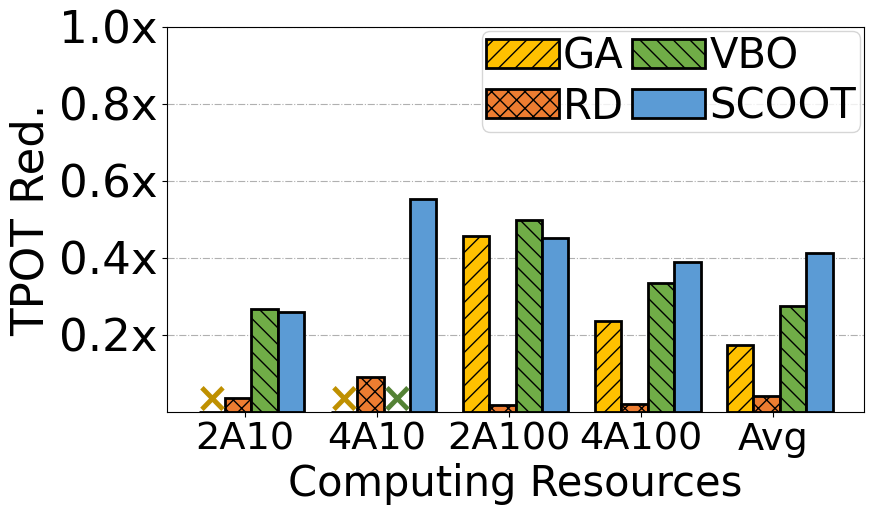}
	\setlength{\abovecaptionskip}{-0.35cm} 
	\caption{TPOT reduction.}
	\label{fig:sql_llama13b_tpot}
\end{subfigure}
\setlength{\abovecaptionskip}{-0.0125cm} 
\caption{SLO optimization for \textbf{SQL} under \textbf{LLAMA2-13B}.}
\label{fig:sql_llama13b}
\vspace{-0.5cm}
\end{figure}

\begin{figure}[t]
\centering
\begin{subfigure}{0.475\linewidth}
	\centering
	\includegraphics[width=1\linewidth]{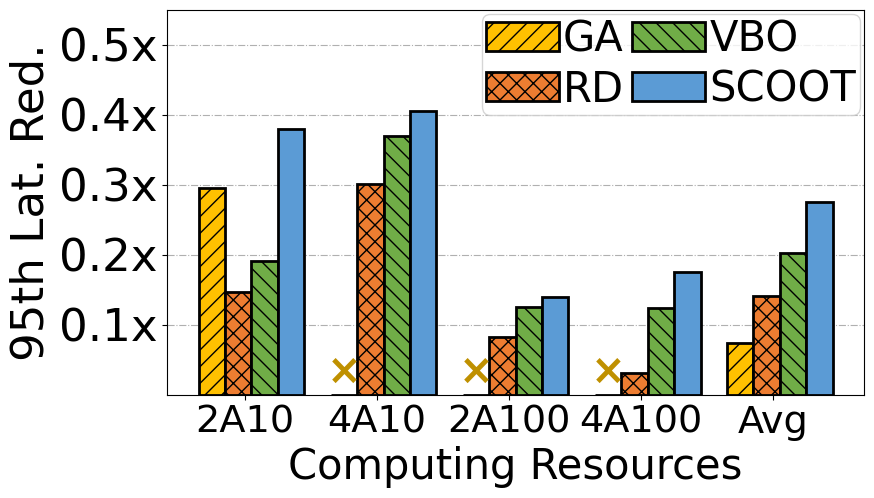}
	\setlength{\abovecaptionskip}{-0.35cm} 
	\caption{\textbf{LLAMA2-7B}.}
	\label{fig:cls_llama7b_latency}
\end{subfigure}
\centering
\begin{subfigure}{0.475\linewidth}
	\centering
	\includegraphics[width=1\linewidth]{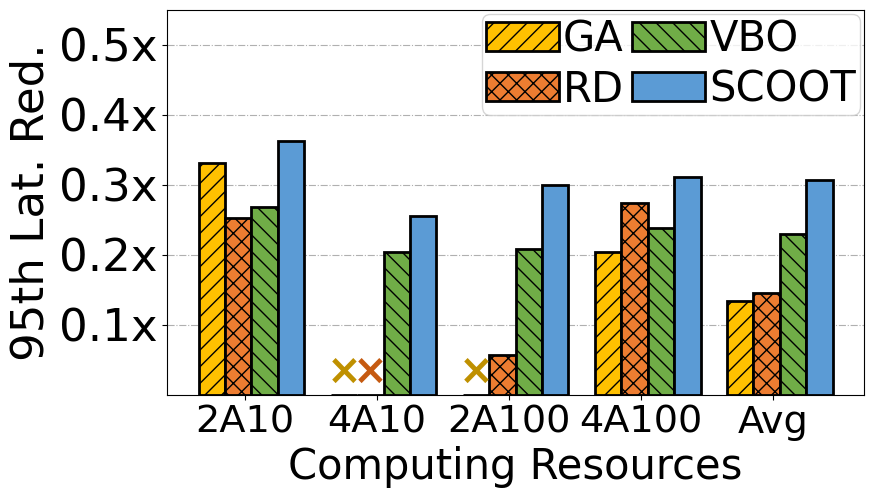}
	\setlength{\abovecaptionskip}{-0.35cm} 
	\caption{\textbf{LLAMA2-13B}.}
	\label{fig:cls_llama13b_latency}
\end{subfigure}
\setlength{\abovecaptionskip}{-0.0125cm} 
\caption{Tail latency reduction for \textbf{CLS}.}
\label{fig:cls_latency}
\vspace{-0.5cm}
\end{figure}

\begin{figure}[t]
\centering
\begin{subfigure}{0.475\linewidth}
	\centering
	\includegraphics[width=1\linewidth]{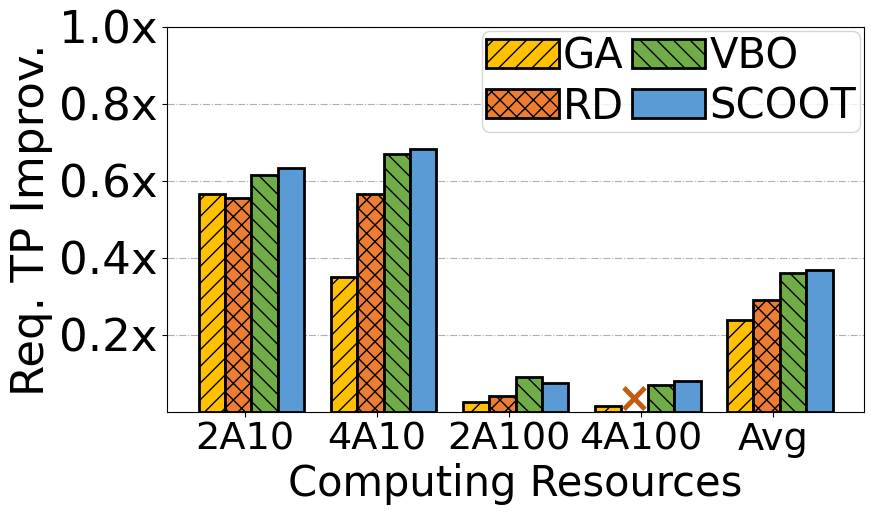}
	\setlength{\abovecaptionskip}{-0.35cm} 
	\caption{\textbf{LLAMA2-7B}.}
	\label{fig:rec_llama7b_tp}
\end{subfigure}
\centering
\begin{subfigure}{0.475\linewidth}
	\centering
	\includegraphics[width=1\linewidth]{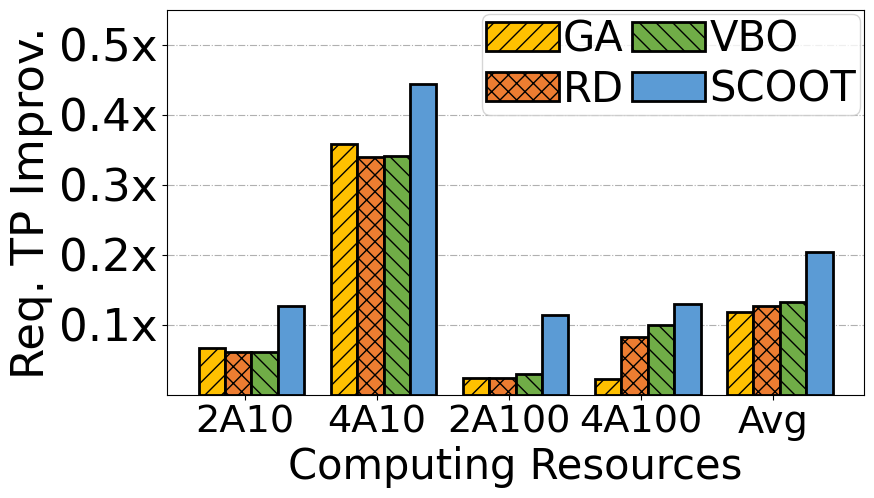}
	\setlength{\abovecaptionskip}{-0.35cm} 
	\caption{\textbf{LLAMA2-13B}.}
	\label{fig:rec_llama13b_tp}
\end{subfigure}
\setlength{\abovecaptionskip}{-0.0125cm} 
\caption{Request TP improvement for \textbf{REC}.}
\label{fig:rec_tp}
\end{figure}

Figures \ref{fig:bot_llama7b}, \ref{fig:bot_llama13b}, \ref{fig:sql_llama7b}, and \ref{fig:sql_llama13b} show that compared to the default configuration and baselines, SCOOT decreases the TTFT and TPOT by up to 98.9\% and 10.5\%  for the BOT application, respectively, and reduces the TTFT and TPOT by up to 99.8\% and 61.0\% for the SQL application, separately. In addition, SCOOT surpasses all the baselines in terms of average TTFT and TPOT reduction across various computing resources for BOT and SQL applications, which confirms SCOOT's superiority in multi-objective optimization.

Figures \ref{fig:cls_latency} and \ref{fig:rec_tp} present that compared to the default configuration and baselines, SCOOT reduces request tail latency by up to 40.6\% for the CLS application and increases request throughput by up to 68.3\% for the REC application. Besides, SCOOT outperforms all the baselines in terms of average tail latency reduction and request throughput improvement across various computing resources for CLS and REC applications, respectively, which confirms SCOOT's superiority in single-objective optimization.

While in rare situations, SCOOT failed to outperform certain baselines in SLO optimization, SCOOT's tuning performance is more consistent. As shown in Fig. \ref{fig:sql_llama13b}, in the case of 2A100, VBO slightly outperforms SCOOT in both TTFT and TPOT reduction. However, in the case of 4A10, VBO fails to find a configuration that is better than the default configuration, while SCOOT greatly reduces TTFT and TPOT. Such a consistently superior performance makes SCOOT outperform all the baselines in the average performance of SLO optimization for various LLMs and applications.

GA optimizes around the initially sampled configurations. With a limited number of observations, both the population size and the number of iterations of GA are restricted to a small value. Thus, GA not only lacks diversity in solutions but also cannot fully utilize historical information to guide the search process, and GA can hardly tackle constraints, hence often performing poorly.

RD is unable to utilize historical information to guide the sampling process and can not handle constraints. However, due to the uniform sampling, RD can explore a large range in the search space and hence has a higher diversity in sampled configurations than GA. Hence, RD often performs better than GA.

By leveraging BO to intelligently utilize historical observations to guide the tuning process, VBO often outperforms RD and GA in SLO optimization. However, by assigning a penalty objective value to unfeasible configurations, the surrogate model will confuse points near infeasible configurations with points that are feasible but perform poorly. This can lead to a lack of subsequent exploration in the adjacent area of infeasible configurations, thus missing out on some potentially superior solutions.

SCOOT’s outstanding performance primarily comes from the employment of BO, which efficiently resolves black-box optimization problems, and the random forest-based POF learning, which intelligently handles hidden constraints without affecting the GP-based surrogate model. Besides, by dynamically adjusting the POF threshold $\Delta$, SCOOT can explore the region near the infeasible configurations after continuous exploration of feasible configurations so as to fully explore the solution space.

\vspace{-0.25cm}
\subsection{Tuning Efficiency}
\vspace{-0.025cm}
We conduct experiments to tune the CSL and BOT applications under 2A100 with different parallelism degrees (PDs). In the experiments, the number of observations is also limited to 30 when PD is set to 2. Figures \ref{fig:batch_cls} and \ref{fig:batch_bot} show the SCOOT's performance in SLO optimization under various PDs, while Fig. \ref{fig:batch_time}  presents the total tuning time under various PDs. 

Figures \ref{fig:batch_cls} and \ref{fig:batch_bot} show that when PD is set to 2, using the parallel suggestion achieves the same performance in SLO optimization as not using it. Besides, as presented in Fig. \ref{fig:batch_time}, when PD is set to 2, the parallel suggestion technique significantly reduces the total tuning time, nearly cutting it in half, which demonstrates that the parallel suggestion can effectively accelerate the tuning process by PD times. Furthermore, Figure \ref{fig:batch_time} shows that compared with the time of configuration evaluation (i.e., stress testing), the time required for configuration suggestion is negligible, which validates the efficiency of applying BO to tune LLM inference engines. 

We also increase PD and conduct experiments. However, we find that when PD is set to a large value, such as greater than 4, SCOOT's performance of SLO optimization is compromised, and increasing the total observation number can mitigate this issue. In the future, we will explore the best number of observations under different PDs to find the most cost-effective acceleration solution. 

\begin{figure}[t]
\centering
\begin{subfigure}{0.332\linewidth}
	\centering
	\includegraphics[width=1\linewidth]{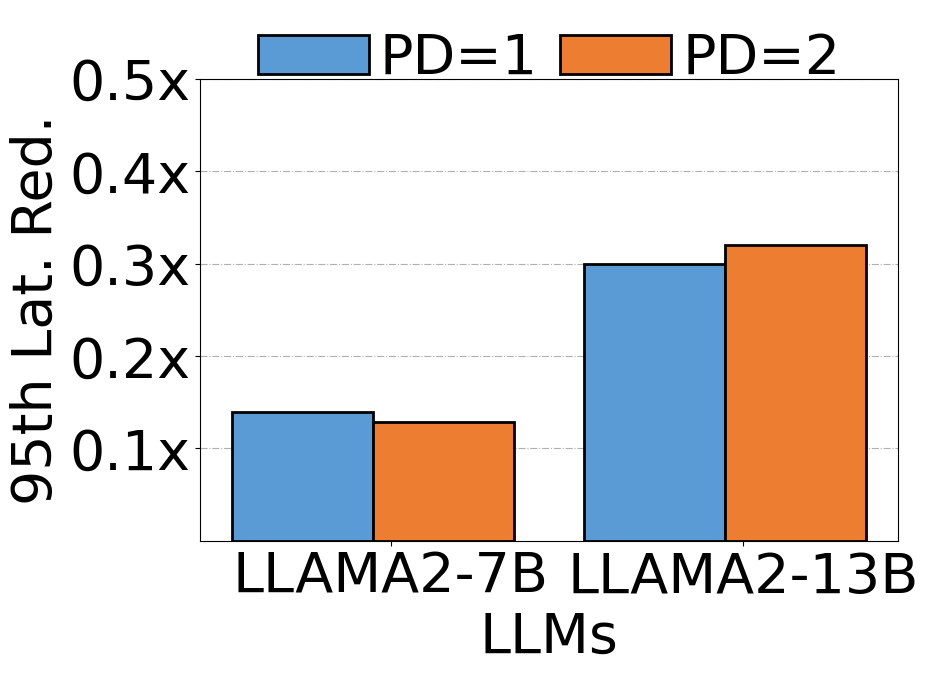}
	\setlength{\abovecaptionskip}{-0.35cm} 
	\caption{Tail latency reduction for \textbf{CLS} under LLAMA2-7B \& 13B.}
	\label{fig:batch_cls}
\end{subfigure}
\centering
\begin{subfigure}{0.325\linewidth}
	\centering
	\includegraphics[width=1\linewidth]{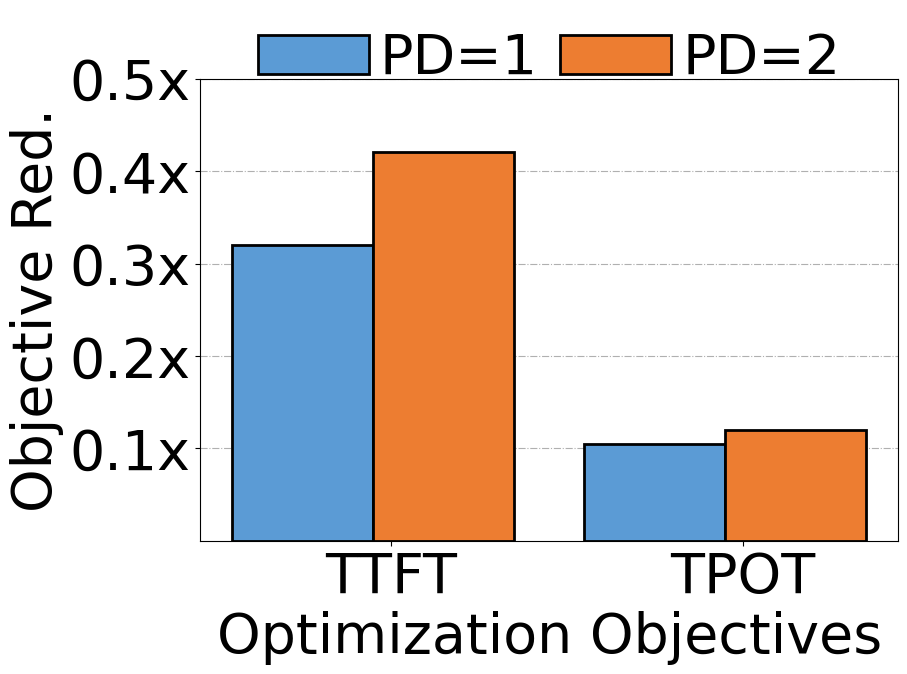}
	\setlength{\abovecaptionskip}{-0.35cm} 
	\caption{TTFT \& TPOT reduction for \textbf{BOT} under LLAMA2-7B.}
	\label{fig:batch_bot}
\end{subfigure}
\begin{subfigure}{0.325\linewidth}
	\centering
	\includegraphics[width=1\linewidth]{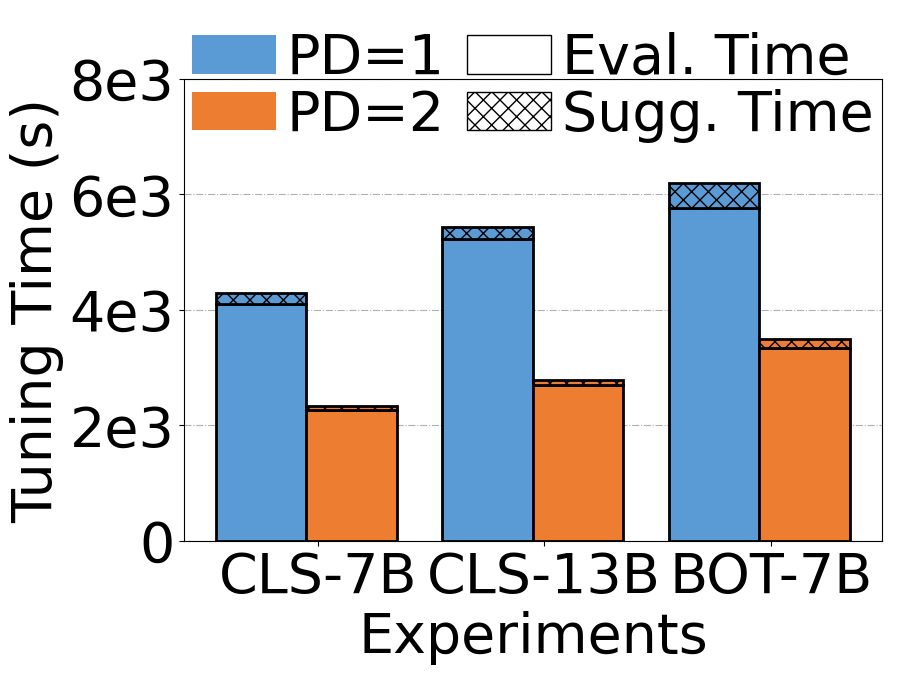}
	\setlength{\abovecaptionskip}{-0.35cm} 
	\caption{Performance tuning time for \textbf{CLS} and \textbf{BOT}.}
	\label{fig:batch_time}
\end{subfigure}
\setlength{\abovecaptionskip}{-0.35cm} 
\caption{Tuning efficiency of the parallel suggestion.}
\label{fig:batch}
\end{figure}

\vspace{-0.25cm}
\subsection{Supplementary Experiments}
\vspace{-0.025cm}
In the appendix, we evaluate the effectiveness of the random forest (RF)-based learning of hidden constraints through ablation studies, confirm the superiority of SCOOT in both the optimality and diversity of SLO optimization in multi-objective situations, and show that SCOOT can outperform baselines and human experts through real-world inference service optimization results, where both vLLM and TensorRT-LLM are leveraged to verify the universality of SCOOT.

\vspace{-0.25cm}
\section{Related Works}
\vspace{-0.025cm}
\label{sec:related_work}
\subsection{Advanced LLM Inference Techniques}
Existing inference engines such as vLLM and TensorRT-LLM support a variety of advanced inference techniques to improve the speed and throughput of LLM inference, including a series of kernel-level optimizations such as continuous batching \cite{yu2022orca}, paged attention \cite{kwon2023efficient}, flash attention \cite{dao2022flashattention, dao2023flashattention2}, and chunked prefill \cite{agrawal2024taming}. Besides, prefix caching \cite{jhaimproving} is employed by inference engines to reuse computed key and value tensors of requests' common prefix texts to reduce redundant memory allocation and computation for newly arrived requests, which can significantly improve the serving efficiency when requests share a long common prefix. Moreover, speculative decoding  \cite{leviathan2023fast, xia2023speculative, miao2024specinfer, kim2024speculative, sun2024spectr} is also adopted by inference engines to accelerate LLM inference, where an efficient draft model is leveraged to generate tokens, and the LLM occasionally refines the draft. 

Some of these technologies are effective in specific scenarios and have been implemented as configurable parameters of inference engines. However, how to combine these technologies to fully exploit the capabilities of inference engines has not been adequately studied, a gap that this study aims to fill.

\vspace{-0.25cm}
\subsection{Automatic Performance Tuning Approaches}
\vspace{-0.05cm}
Performance tuning techniques have been widely applied across a wide variety of fields, including database tuning \cite{zhang2021restune, lao2023gptuner, yang2024vdtuner}, Spark configuration tuning  \cite{gounaris2017dynamic, fekry2020tune, li2023towards, shen2023rover}, compiler optimization \cite{hellsten2023baco}, and tuning of web-relevant applications \cite{dalibard2017boat, li2018bayesian}. The vast majority of these performance tuning studies adopt BO to find optimized parameter configurations since BO is theoretically grounded and can efficiently learn the relationship between performance and parameters from evaluations, thus intelligently tuning parameters for performance improvement. Because of these advantages, BO is also employed in resource allocation studies \cite{patel2020clite,roy2021satori, yang2023cotuner}.

To the best of our knowledge, previous studies have not focused on tuning LLM inference engines and can not address all the three unique challenges faced in inference engine tuning.

\vspace{-0.2cm}
\section{Conclusion}
\vspace{-0.05cm}
In this paper, we propose SCOOT to optimize SLOs for LLM inference services via BO-based automatic inference engine tuning. SCOOT uses RF to learn hidden constraints and employs the parallel suggestion to speed up tuning. Extensive experiments are conducted to confirm both the effectiveness and efficiency of SCOOT.

\vspace{-0.05cm}
\begin{acks}
This work is supported in part by the Postgraduate Research \& Practice Innovation Program of Jiangsu Province (KYCX24\_0247), the Nanjing Key S\&T Special Projects (202309006), and the Ant Research Program of Ant Group.
\end{acks}

\bibliographystyle{unsrt}
\balance
\bibliography{reference}

\begin{thebibliography}{10}

\bibitem{min2023recent}
Bonan Min, Hayley Ross, Elior Sulem, et~al.
\newblock Recent advances in natural language processing via large pre-trained
  language models: A survey.
\newblock {\em ACM Computing Surveys}, 56(2):1--40, 2023.

\bibitem{alibaba_bailian}
Alibaba cloud bailian platform.
\newblock \url{https://www.aliyun.com/product/bailian}, 2024.

\bibitem{aws_sagemaker}
Aws sagemaker.
\newblock \url{https://aws.amazon.com/sagemaker}, 2024.

\bibitem{vllm}
vllm.
\newblock \url{https://github.com/vllm-project/vllm}, 2024.

\bibitem{tensorrt_llm}
Tensorrt-llm.
\newblock \url{https://github.com/NVIDIA/TensorRT-LLM}, 2024.

\bibitem{yu2022orca}
Gyeong-In Yu, Joo~Seong Jeong, Geon-Woo Kim, et~al.
\newblock Orca: A distributed serving system for transformer-based generative
  models.
\newblock In {\em 16th USENIX Symposium on Operating Systems Design and
  Implementation (OSDI 22)}, pages 521--538, 2022.

\bibitem{kwon2023efficient}
Woosuk Kwon, Zhuohan Li, Siyuan Zhuang, et~al.
\newblock Efficient memory management for large language model serving with
  pagedattention.
\newblock In {\em Proceedings of the 29th Symposium on Operating Systems
  Principles}, pages 611--626, 2023.

\bibitem{agrawal2024taming}
Amey Agrawal, Nitin Kedia, Ashish Panwar, Jayashree Mohan, et~al.
\newblock Taming throughput-latency tradeoff in llm inference with
  sarathi-serve.
\newblock In {\em 18th USENIX Symposium on Operating Systems Design and
  Implementation (OSDI 24)}, pages 117--134, 2024.

\bibitem{xu2024deploying}
Wenchao Xu, Jinyu Chen, Peirong Zheng, et~al.
\newblock Deploying foundation model powered agent services: A survey.
\newblock 2024.

\bibitem{herodotou2020survey}
Herodotos Herodotou, Yuxing Chen, and Jiaheng Lu.
\newblock A survey on automatic parameter tuning for big data processing
  systems.
\newblock {\em ACM Computing Surveys (CSUR)}, 53(2):1--37, 2020.

\bibitem{homem2014monte}
Tito Homem-de Mello and G{\"u}zin Bayraksan.
\newblock Monte carlo sampling-based methods for stochastic optimization.
\newblock {\em Surveys in Operations Research and Management Science},
  19(1):56--85, 2014.

\bibitem{katoch2021review}
Sourabh Katoch, Sumit~Singh Chauhan, and Vijay Kumar.
\newblock A review on genetic algorithm: past, present, and future.
\newblock {\em Multimedia tools and applications}, 80:8091--8126, 2021.

\bibitem{padakandla2021survey}
Sindhu Padakandla.
\newblock A survey of reinforcement learning algorithms for dynamically varying
  environments.
\newblock {\em ACM Computing Surveys (CSUR)}, 54(6):1--25, 2021.

\bibitem{wang2023recent}
Xilu Wang, Yaochu Jin, Sebastian Schmitt, and Markus Olhofer.
\newblock Recent advances in bayesian optimization.
\newblock {\em ACM Computing Surveys}, 55(13s):1--36, 2023.

\bibitem{khan2002multi}
Nazan Khan, David~E Goldberg, and Martin Pelikan.
\newblock Multi-objective bayesian optimization algorithm.
\newblock In {\em Proceedings of the 4th Annual Conference on Genetic and
  Evolutionary Computation}, pages 684--684, 2002.

\bibitem{yang2019multi}
Kaifeng Yang, Michael Emmerich, Andr{\'e} Deutz, and Thomas B{\"a}ck.
\newblock Multi-objective bayesian global optimization using expected
  hypervolume improvement gradient.
\newblock {\em Swarm and evolutionary computation}, 44:945--956, 2019.

\bibitem{daulton2020differentiable}
Samuel Daulton, Maximilian Balandat, and Eytan Bakshy.
\newblock Differentiable expected hypervolume improvement for parallel
  multi-objective bayesian optimization.
\newblock {\em Advances in Neural Information Processing Systems},
  33:9851--9864, 2020.

\bibitem{leviathan2023fast}
Yaniv Leviathan, Matan Kalman, and Yossi Matias.
\newblock Fast inference from transformers via speculative decoding.
\newblock In {\em International Conference on Machine Learning}, pages
  19274--19286. PMLR, 2023.

\bibitem{caflisch1998monte}
Russel~E Caflisch.
\newblock Monte carlo and quasi-monte carlo methods.
\newblock {\em Acta numerica}, 7:1--49, 1998.

\bibitem{snoek2014input}
Jasper Snoek, Kevin Swersky, Rich Zemel, and Ryan Adams.
\newblock Input warping for bayesian optimization of non-stationary functions.
\newblock In {\em International conference on machine learning}, pages
  1674--1682. PMLR, 2014.

\bibitem{seeger2004gaussian}
Matthias Seeger.
\newblock Gaussian processes for machine learning.
\newblock {\em International journal of neural systems}, 14(02):69--106, 2004.

\bibitem{cowen2022hebo}
Alexander~I Cowen-Rivers, Wenlong Lyu, Rasul Tutunov, et~al.
\newblock Hebo: Pushing the limits of sample-efficient hyper-parameter
  optimisation.
\newblock {\em Journal of Artificial Intelligence Research}, 74:1269--1349,
  2022.

\bibitem{zhang2021efficient}
Shuhan Zhang, Fan Yang, Changhao Yan, et~al.
\newblock An efficient batch-constrained bayesian optimization approach for
  analog circuit synthesis via multiobjective acquisition ensemble.
\newblock {\em IEEE Transactions on Computer-Aided Design of Integrated
  Circuits and Systems}, 41(1):1--14, 2021.

\bibitem{srinivas2010gaussian}
Niranjan Srinivas, Andreas Krause, Sham Kakade, and Matthias Seeger.
\newblock Gaussian process optimization in the bandit setting: no regret and
  experimental design.
\newblock In {\em Proceedings of the 27th International Conference on Machine
  Learning}, pages 1015--1022, 2010.

\bibitem{hellsten2023baco}
Erik~Orm Hellsten, Artur Souza, Johannes Lenfers, et~al.
\newblock Baco: A fast and portable bayesian compiler optimization framework.
\newblock In {\em Proceedings of the 28th ACM International Conference on
  Architectural Support for Programming Languages and Operating Systems, Volume
  4}, pages 19--42, 2023.

\bibitem{pedregosa2011scikit}
Fabian Pedregosa, Ga{\"e}l Varoquaux, Alexandre Gramfort, et~al.
\newblock Scikit-learn: Machine learning in python.
\newblock {\em the Journal of machine Learning research}, 12:2825--2830, 2011.

\bibitem{llama2}
Hugo Touvron, Louis Martin, Kevin Stone, et~al.
\newblock Llama 2: Open foundation and fine-tuned chat models.
\newblock 2023.

\bibitem{dao2022flashattention}
Tri Dao, Dan Fu, Stefano Ermon, et~al.
\newblock Flashattention: Fast and memory-efficient exact attention with
  io-awareness.
\newblock {\em Advances in Neural Information Processing Systems},
  35:16344--16359, 2022.

\bibitem{dao2023flashattention2}
Tri Dao.
\newblock Flashattention-2: Faster attention with better parallelism and work
  partitioning.
\newblock In {\em The Twelfth International Conference on Learning
  Representations}, 2023.

\bibitem{jhaimproving}
Nikhil Jha and Kevin Wang.
\newblock Improving large language model throughput with efficient long-term
  memory management.

\bibitem{xia2023speculative}
Heming Xia, Tao Ge, Peiyi Wang, et~al.
\newblock Speculative decoding: Exploiting speculative execution for
  accelerating seq2seq generation.
\newblock In {\em Findings of the Association for Computational Linguistics:
  EMNLP 2023}, pages 3909--3925, 2023.

\bibitem{miao2024specinfer}
Xupeng Miao, Gabriele Oliaro, Zhihao Zhang, et~al.
\newblock Specinfer: Accelerating large language model serving with tree-based
  speculative inference and verification.
\newblock In {\em Proceedings of the 29th ACM International Conference on
  Architectural Support for Programming Languages and Operating Systems, Volume
  3}, pages 932--949, 2024.

\bibitem{kim2024speculative}
Sehoon Kim, Karttikeya Mangalam, Suhong Moon, et~al.
\newblock Speculative decoding with big little decoder.
\newblock {\em Advances in Neural Information Processing Systems}, 36, 2024.

\bibitem{sun2024spectr}
Ziteng Sun, Ananda~Theertha Suresh, Jae~Hun Ro, et~al.
\newblock Spectr: Fast speculative decoding via optimal transport.
\newblock {\em Advances in Neural Information Processing Systems}, 36, 2024.

\bibitem{zhang2021restune}
Xinyi Zhang, Hong Wu, Zhuo Chang, et~al.
\newblock Restune: Resource oriented tuning boosted by meta-learning for cloud
  databases.
\newblock In {\em Proceedings of the 2021 international conference on
  management of data}, pages 2102--2114, 2021.

\bibitem{lao2023gptuner}
Jiale Lao, Yibo Wang, Yufei Li, et~al.
\newblock Gptuner: A manual-reading database tuning system via gpt-guided
  bayesian optimization.
\newblock 2023.

\bibitem{yang2024vdtuner}
Tiannuo Yang, Wen Hu, Wangqi Peng, et~al.
\newblock Vdtuner: Automated performance tuning for vector data management
  systems.
\newblock In {\em 2024 IEEE 40th International Conference on Data Engineering
  (ICDE)}, pages 4357--4369, 2024.

\bibitem{gounaris2017dynamic}
Anastasios Gounaris, Georgia Kougka, and Ruben~others Tous.
\newblock Dynamic configuration of partitioning in spark applications.
\newblock {\em IEEE Transactions on Parallel and Distributed Systems},
  28(7):1891--1904, 2017.

\bibitem{fekry2020tune}
Ayat Fekry, Lucian Carata, Thomas Pasquier, et~al.
\newblock To tune or not to tune? in search of optimal configurations for data
  analytics.
\newblock In {\em Proceedings of the 26th ACM SIGKDD International Conference
  on Knowledge Discovery \& Data Mining}, pages 2494--2504, 2020.

\bibitem{li2023towards}
Yang Li, Huaijun Jiang, Yu~Shen, et~al.
\newblock Towards general and efficient online tuning for spark.
\newblock {\em Proceedings of the VLDB Endowment}, 16(12):3570--3583, 2023.

\bibitem{shen2023rover}
Yu~Shen, Xinyuyang Ren, Yupeng Lu, Huaijun Jiang, Huanyong Xu, Di~Peng, Yang
  Li, Wentao Zhang, and Bin Cui.
\newblock Rover: An online spark sql tuning service via generalized transfer
  learning.
\newblock In {\em Proceedings of the 29th ACM SIGKDD Conference on Knowledge
  Discovery and Data Mining}, pages 4800--4812, 2023.

\bibitem{dalibard2017boat}
Valentin Dalibard, Michael Schaarschmidt, and Eiko Yoneki.
\newblock Boat: Building auto-tuners with structured bayesian optimization.
\newblock In {\em Proceedings of the 26th International Conference on World
  Wide Web}, pages 479--488, 2017.

\bibitem{li2018bayesian}
Dan Li and Evangelos Kanoulas.
\newblock Bayesian optimization for optimizing retrieval systems.
\newblock In {\em Proceedings of the Eleventh ACM International Conference on
  Web Search and Data Mining}, pages 360--368, 2018.

\bibitem{patel2020clite}
Tirthak Patel and Devesh Tiwari.
\newblock Clite: Efficient and qos-aware co-location of multiple
  latency-critical jobs for warehouse scale computers.
\newblock In {\em 2020 IEEE International Symposium on High Performance
  Computer Architecture (HPCA)}, pages 193--206. IEEE, 2020.

\bibitem{roy2021satori}
Rohan~Basu Roy, Tirthak Patel, and Devesh Tiwari.
\newblock Satori: efficient and fair resource partitioning by sacrificing
  short-term benefits for long-term gains.
\newblock In {\em 2021 ACM/IEEE 48th Annual International Symposium on Computer
  Architecture (ISCA)}, pages 292--305. IEEE, 2021.

\bibitem{yang2023cotuner}
Tiannuo Yang, Ruobing Chen, Yusen Li, Xiaoguang Liu, and Gang Wang.
\newblock Cotuner: A hierarchical learning framework for coordinately
  optimizing resource partitioning and parameter tuning.
\newblock In {\em Proceedings of the 52nd International Conference on Parallel
  Processing}, pages 317--326, 2023.

\bibitem{stelmack1998genetic}
Marc Stelmack, Nari Nakashima, and Stephen Batill.
\newblock Genetic algorithms for mixed discrete/continuous optimization in
  multidisciplinary design.
\newblock In {\em 7th AIAA/USAF/NASA/ISSMO Symposium on Multidisciplinary
  Analysis and Optimization}, page 4771, 1998.

\bibitem{deb2002fast}
Kalyanmoy Deb, Amrit Pratap, Sameer Agarwal, and TAMT Meyarivan.
\newblock A fast and elitist multiobjective genetic algorithm: Nsga-ii.
\newblock {\em IEEE transactions on evolutionary computation}, 6(2):182--197,
  2002.

\bibitem{llama.cpp}
llama.cpp: Llm inference in c/c++.
\newblock \url{https://github.com/ggerganov/llama.cpp}, 2024.

\bibitem{vllm_doc}
vllm performance and tuning.
\newblock \url{https://docs.vllm.ai/en/v0.4.2/models/performance.html}, 2024.

\bibitem{tensorrt_doc}
Best practices for tuning the performance of tensorrt-llm.
\newblock
  \url{https://github.com/NVIDIA/TensorRT-LLM/blob/main/docs/source/performance/perf-best-practices.md},
  2024.

\end{thebibliography}

\appendix
\section{Appendix}

\subsection{Conflicting TTFT and TPOT}
\label{sec:conflict}
\begin{figure}[t]
\centering
\begin{subfigure}{0.485\linewidth}
	\centering
	\includegraphics[width=1\linewidth]{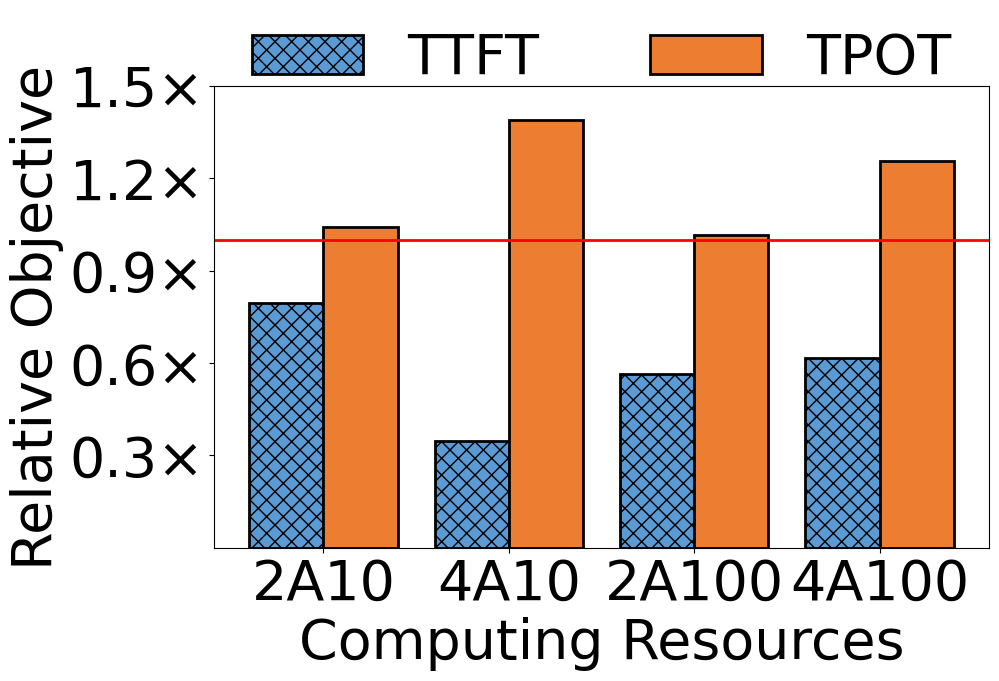}
	\setlength{\abovecaptionskip}{-0.35cm} 
	\caption{TTFT-only optimization.}
	\label{fig:mtv_conflict_ttft}
\end{subfigure}
\centering
\begin{subfigure}{0.485\linewidth}
	\centering
	\includegraphics[width=1\linewidth]{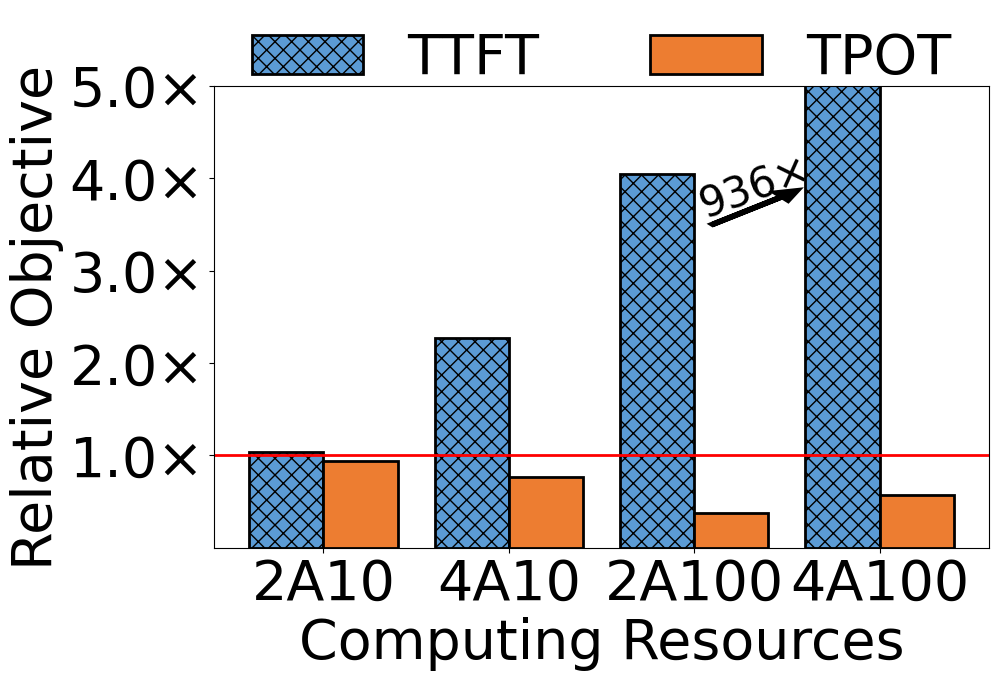}
	\setlength{\abovecaptionskip}{-0.35cm} 
	\caption{TPOT-only optimization.}
	\label{fig:mtv_conflict_tpot}
\end{subfigure}
\setlength{\abovecaptionskip}{-0.01cm} 
\caption{Conflicting TTFT and TOPT. Decreased TTFT often leads to increased TPOT, and vice versa. The red line indicates the objective value under the default parameter configuration.}
\vspace{-0.05cm}
\label{fig:mtv_conflict}
\end{figure}

\begin{figure}[t]
\centering
\begin{subfigure}{0.475\linewidth}
	\centering
	\includegraphics[width=1\linewidth]{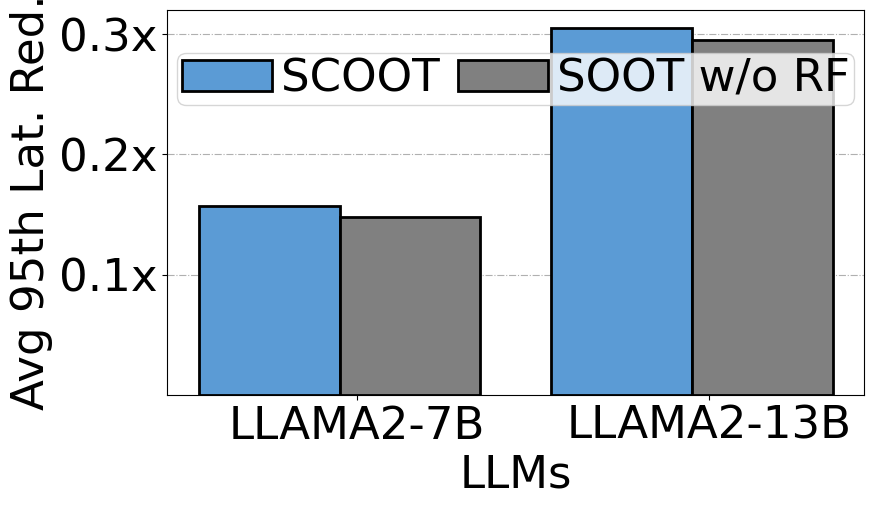}
	\setlength{\abovecaptionskip}{-0.35cm} 
	\caption{Tail latency reduction for \textbf{CLS} under LLAMA2-7B \& 13B.}
	\label{fig:abla_cls}
\end{subfigure}
\centering
\begin{subfigure}{0.475\linewidth}
	\centering
	\includegraphics[width=1\linewidth]{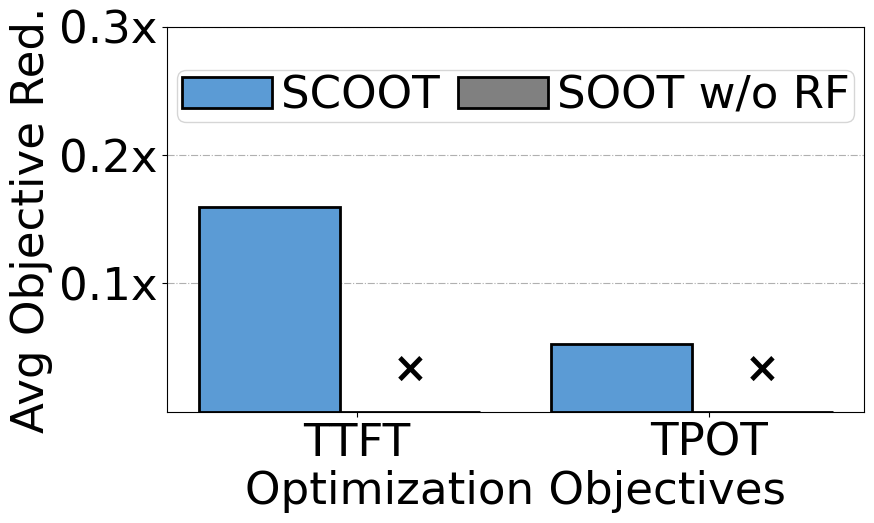}
	\setlength{\abovecaptionskip}{-0.35cm} 
	\caption{TTFT \& TPOT reduction for \textbf{BOT} under LLAMA2-7B.}
	\label{fig:abla_bot}
\end{subfigure}
\setlength{\abovecaptionskip}{-0.01cm} 
\caption{Effectiveness of random forest-based POF learning.}
\label{fig:abla}
\end{figure}

We separately take TTFT and TPOT as the only optimization objectives and conduct grid searches for inference services of LLAMA2-13B deployed on various computing resources under the BOT request trace. Experimental results are shown in Fig. \ref{fig:mtv_conflict}, which confirms that for TTFT and TPOT, optimizing for one always compromised the other. 
The reason is that, to decrease TPOT, it is significant to decrease the batch size controlled by the parameter \verb|max-num-seqs| to reduce the per-token latency, which causes a long queuing time for requests, thus leading to a large TTFT. Conversely, to reduce TTFT, a large batch size is always set to allow newly arrived requests to be processed promptly without queuing for a long time. However, a large batch size often slows down inference, leading to increased TPOT.

\vspace{-0.25cm}
\subsection{Known Constraints of Inference Engines}
\vspace{-0.025cm}
\label{sec:known_constraints}
\vspace{-0.125cm}
\subsubsection{\textbf{vLLM}} For vLLM-0.4.2 and vLLM-0.5.5, the known constraints include that \verb|max-num-batched-tokens| must be greater than or equal to \verb|max_num_seqs|, the \verb|enable-chunked-prefill| and \verb|enable-prefix-caching| cannot be set to \verb|True| simultaneously, and when the \verb|enable-chunked-prefill| is not enabled, \verb|max-num-batched-tokens| needs to be greater than or equal to the max model length specified in the model configuration file. 
\vspace{-0.125cm}
\subsubsection{\textbf{TensorRT-LLM}} For TensorRT-LLM-0.15.5, the known constraints include that \verb|max-num-tokens| must be greater than or equal to \verb|max_batch_size|, \verb|use-paged-context-fmha| must be set to \verb|True| if \verb|enable-chunked-context| is enabled, and when the \verb|enable-chunked-contex| is not enabled, \verb|max-num-tokens| needs to be greater than or equal to the max model length specified in the model configuration file.

\vspace{-0.25cm}
\subsection{Baseline Description}
\vspace{-0.025cm}
\label{sec:baseline}
\begin{itemize}
\item \textbf{Random Sampling (RD)}: Sobol sequence-based Quasi-Monte Carlo \cite{caflisch1998monte} is utilized to uniformly sample configuration points from the search space.
\item \textbf{Genetic Algorithm (GA)}: Mixed variable GA \cite{stelmack1998genetic} is used to resolve single-objective optimization problems while NSGA2 \cite{deb2002fast} is utilized to handle multi-objective situations. For GA, we assign a penalty objective value when constraints are violated and set the population size to 10.
\item \textbf{Vanilla BO (VBO)}: VBO uses UCB and EHVI as acquisition functions to handle single-objective and multi-objective optimization situations, respectively. For VBO, we assign a penalty objective value when constraints are violated and use a random forest-based surrogate model to learn objective functions and constraints at the same time. 
\end{itemize}
All of these approaches are developed using HEBO \cite{cowen2022hebo}, which has highly optimized implementations. Therefore, we think they are competitive and suitable as baselines.

\vspace{-0.25cm}
\subsection{Ablation Study Experiments}
\vspace{-0.025cm}
We conduct experiments to confirm the effectiveness of the random forest (RF)-based learning of hidden constraints under 2A100 and 4A100. The average SLO optimization results across various computing resources are presented in Fig. \ref{fig:abla}. Since requests from the CLS application seldom cause engine crashes, not applying RF-based POF learning has little impact on SCOOT's performance. However, since BOT application requests cause many hidden constraints, RF-based POF learning can greatly enhance SLO optimization.

\begin{figure}[H]
\centering
\begin{subfigure}{0.493\linewidth}
	\centering
	\includegraphics[width=1\linewidth]{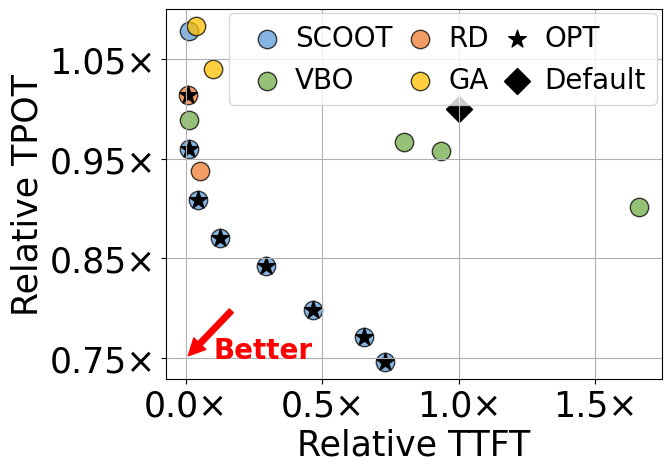}
	\setlength{\abovecaptionskip}{-0.35cm} 
	\caption{Optimized TTFT \& TPOT under \textbf{LLAMA2-7B}.}
	\label{fig:diversity_bot_llama7b}
\end{subfigure}
\centering
\begin{subfigure}{0.493\linewidth}
	\centering
	\includegraphics[width=1\linewidth]{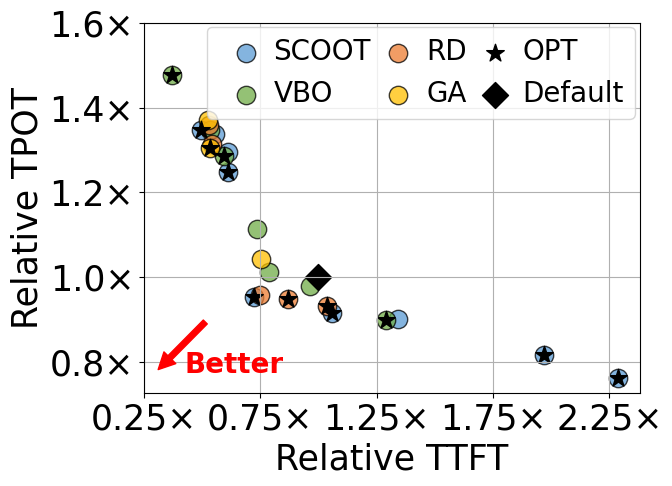}
	\setlength{\abovecaptionskip}{-0.35cm} 
	\caption{Optimized TTFT \& TPOT  under \textbf{LLAMA2-13B}.}
	\label{fig:diversity_bot_llama13b}
\end{subfigure}
\setlength{\abovecaptionskip}{-0.35cm} 
\caption{Solution quality for \textbf{BOT}.}
\vspace{-0.75cm}
\label{fig:bot_diversity}
\end{figure}
\begin{figure}[H]
\centering
\begin{subfigure}{0.49\linewidth}
	\centering
	\includegraphics[width=1\linewidth]{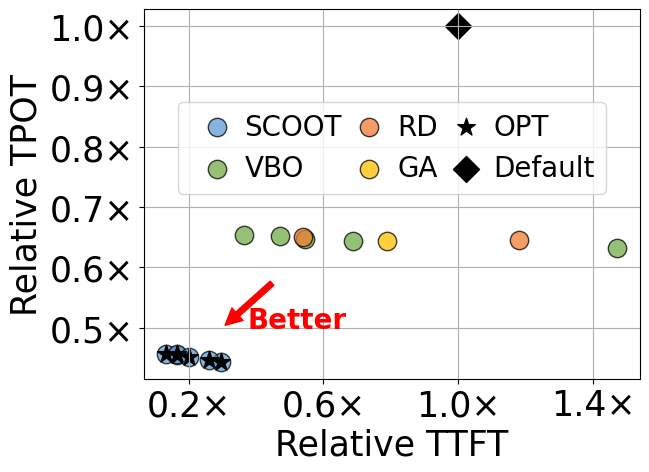}
	\setlength{\abovecaptionskip}{-0.35cm} 
	\caption{Optimized TTFT \& TPOT  under \textbf{LLAMA2-7B}.}
	\label{fig:diversity_sql_llama7b}
\end{subfigure}
\centering
\begin{subfigure}{0.49\linewidth}
	\centering
	\includegraphics[width=1\linewidth]{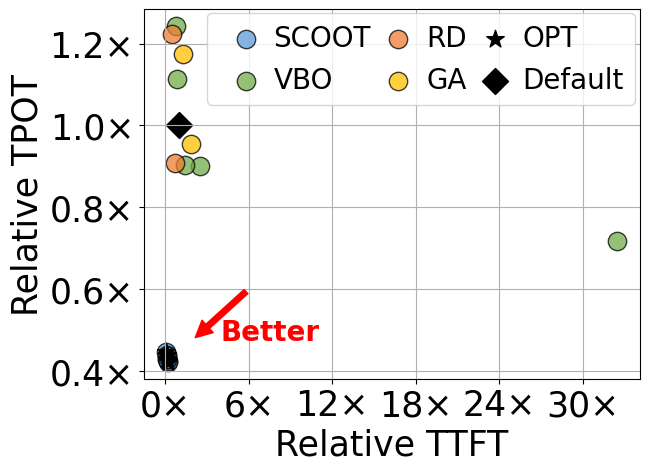}
	\setlength{\abovecaptionskip}{-0.35cm} 
	\caption{Optimized TTFT \& TPOT  under \textbf{LLAMA2-13B}.}
	\label{fig:diversity_sql_llama13b}
\end{subfigure}
\setlength{\abovecaptionskip}{-0.05cm} 
\caption{Solution quality for \textbf{SQL}.}
\vspace{-0.25cm}
\label{fig:sql_diversity}
\end{figure}

\subsection{Solution Quality Experiments}
\label{sec:solution_quality}
For multi-objective optimization, we evaluate the quality of solutions from two perspectives. First, the optimality of solutions, that is, whether the optimized parameter configurations lie on the Pareto frontier of TTFT and TPOT. Second, we focus on the diversity of solutions, that is, whether the range of optimized parameter configurations on the Pareto frontier is sufficiently broad to offer customers promising optional configurations to select based on their TTFT and TPOT requirements. 

Figures \ref{fig:bot_diversity} and \ref{fig:sql_diversity} plot the Pareto sets of configuration points sampled in the process of tuning the BOT and SQL applications under 4A10 for SCOOT and baselines. The plotted points represent the ``locally optimal solutions" identified by each tuning method. Additionally, we mark the global Pareto set from these locally optimal configuration points with $\boldsymbol{\star}$. From Fig. \ref{fig:bot_diversity}, we can observe that for BOT, the majority of globally optimal configuration points are contributed by SCOOT. Besides, SCOOT yields a wider range of its globally optimal points compared with baselines. From Fig. \ref{fig:sql_diversity}, we can find that for SQL, all the globally optimal configuration points are contributed by SCOOT. These experimental results confirm the superiority of SCOOT in both the optimality and diversity of SLO optimization in multi-objective situations.

SCOOT achieves a broader range of solutions because it uses EHVI as the acquisition function for multi-objective problems to optimize the Pareto front of TTFT and TPOT. Besides, since the default configuration is chosen as the reference point, EHVI ensures that the TTFT and TPOT of the configuration with maximal HV are both better than the default, achieving the goal of optimizing both objectives simultaneously. Since VBO also employs EHVI, its solution range is broad. However, because it cannot effectively handle constraints, the optimality of the solutions is compromised.

\vspace{-0.2cm}
\subsection{Real-World Performance Experiments}
\label{sec:appendix_real_world_performance}
Two real-world LLM inference services at Ant Group aim to maximize throughput while maintaining acceptable TTFT and TPOT. Both services are applications for short story generation, with average input lengths of 1,000 and 500 tokens, respectively, and an average output length of 300 tokens for both. We refer to these two inference services as story-teller-long (STL) and story-teller-short (STS) separately. Both services deploy an LLM with 80 billion parameters on 8 NVIDIA L20 48GB GPUs. During the online serving of STS and STL, the client uses concurrency to limit the maximum number of requests the service handles simultaneously, and we also include the concurrency as a parameter for optimization.

We adopt vLLM-0.5.5 and TensorRT-LLM-0.15.0 as inference engines, respectively, and conduct experiments to optimize throughput for these two inference services. The TensorRT-LLM parameters to be tuned are listed in Table \ref{tab:tune_param_trt}.  We have compared the tuning results of SCOOT with the default parameters of the inference engine, the recommended parameters from human algorithm experts at Ant Group, and VBO. Since VBO has been previously demonstrated to always outperform both GA and RD strategies in our experiments, we have not included comparisons with RD and GA. We filter out parameter configurations that do not meet the requirements for TTFT and TPOT and select the configurations with the highest throughput. The experimental results are presented in Table \ref{tab:real_world}, where the throughput is measured in requests per second (RPS).

Human experts always simply enable certain techniques, which leads to sub-optimal results. Both STL and STS requests have a short common prefix. For vLLM-0.5.5, since chunked prefill and prefix caching are incompatible, human experts enable chunked prefill for STL and prefix caching for STS, respectively. For TensorRT-LLM 0.15.0, human experts enable both chunked prefill and prefix caching for STL and STS. However, due to the lack of sophisticated tuning of the client concurrency and parameters such as \verb|tokens-per-block| and \verb|max-batch-size|, the throughput improvement is always limited. From the experimental results, we can see that SCOOT consistently improves the throughput of two real-world inference services under various inference engines and significantly outperforms all baselines, which confirms the universality and practicality of SCOOT. 

\vspace{-0.25cm}
\begin{table}[H]
\centering
\setlength{\abovecaptionskip}{-0.001cm} 
\caption{TensorRT-LLM PARAMETERS TO TUNE.}
\begin{tabular}{c|c|c}
	\hline
	Configuration Parameter & Type   	& Range  	                  \\ \hline\hline
	\verb|tp-size|          & Int			& [1, \#GPUs]       	           	            \\ \hline
	\verb|max-batch-size|          & Int			& [64, 8192]       	         	            \\ \hline
	\verb|max-num-tokens|		   & Int   & [64, 8192]       	      				\\ \hline
	\verb|tokens-per-block|          & Enum		& \{\verb|4|, \verb|8|, \verb|16|, \verb|32|, \verb|64|, \verb|128|\}	 				\\ \hline
	\verb|use-paged-context-fmha|		   & Bool			& \{\verb|True|, \verb|False|\}			 			\\ \hline
	\verb|enable-chunked-context|		   & Bool			& \{\verb|True|, \verb|False|\}						\\ \hline
	\verb|enable-block-reuse|		   &  Bool		& \{\verb|True|, \verb|False|\}			 			\\ \hline
	\verb|capacity-scheduler-policy|		   &  Enum			& \{\verb|FCFS|, \verb|EQUAL|\}			 			\\ \hline
	\verb|context-chunking-policy|		   &  Enum		& \{\verb|MAX_UTI|, \verb|NO_EVICT|\}		 			\\ \hline		            
\end{tabular}
\vspace{-0.5cm}
\label{tab:tune_param_trt}
\end{table} 

\begin{table}[H]
\setlength{\abovecaptionskip}{-0.001cm} 
\caption{THROUGHPUT OPTIMIZATION RESULTS FOR REAL-WORLD INFERENCE SERVICES.}
\begin{tabular}{c|cc|cc}
	\hline
	& \multicolumn{2}{c|}{vLLM} & \multicolumn{2}{c}{TensorRT-LLM} \\ \cline{2-5} 
	Method  & STL         & STS         & STL             & STS            \\ \hline\hline
	Default & 0.527          & 0.948      & 0.557              & 0.985             \\ \hline
	Expert  & 0.528          & 1.004         & 0.598             & 1.039           \\ \hline
	VBO     & 0.595          & 1.010          & 0.670              & 1.180             \\ \hline
	SCOOT   & \textbf{0.695}          & \textbf{1.211}         & \textbf{0.763}              & \textbf{1.300}            \\ \hline
\end{tabular}
\vspace{-0.25cm}
\label{tab:real_world}
\end{table}

\subsection{Discussion}
\label{sec:appendix_discussion}
\subsubsection{\textbf{Application Scope}}
SCOOT only requires the LLM inference engine to provide tunable parameters and an API-enabled entry point (e.g., FastAPI or OpenAI server). Since both widely-used LLM inference engines in cloud data centers (e.g., vLLM and TensorRT-LLM) and those deployed on personal devices with constraint resources (e.g., llama.cpp \cite{llama.cpp}) meet these requirements, SCOOT can be universally applied across a wide range of LLM deployment scenarios.
\vspace{-0.05cm}
\subsubsection{\textbf{Parameter Selection}} The selection of parameters and their value ranges significantly impact the tuning results. We select the parameters that are generally applicable across most scenarios and are of primary concern to customers, as indicated in the official documentation of vLLM \cite{vllm_doc} and TensorRT-LLM \cite{tensorrt_doc}. Furthermore, to ensure a more comprehensive optimization process, we intentionally include broad value ranges of the configuration parameters for performance tuning.
\vspace{-0.05cm} 
\subsubsection{\textbf{Final Deployed Product}} We leverage existing company infrastructure and develop supplemental modules to ensure smooth deployment of SCOOT. Specifically, we implement several additional modules around SCOOT, including a dataset generation module to extract the request traces from the logging system into our database and utilize scripts to convert them into SCOOT-compatible formats, and a configuration parser to parse user-defined tuning parameters, including parameter names and their ranges. If parameters-to-tune are not specified by users, a default set of parameters and ranges will also be made available to all users. Besides, a post-processing module is developed to store the suggested parameters in the database and present them to customers for selection after SCOOT finishes parameter tuning.

\end{document}